\documentclass [11pt]{article}

\usepackage{natbib}
\usepackage {graphicx}
\usepackage{epsfig,amsmath,latexsym,amssymb}
\usepackage{bm}

\usepackage{booktabs}
\usepackage{multirow}
\usepackage{caption}

\renewcommand{\baselinestretch}{1.3}

\newtheorem{remark}{Remark}[section]

\begin{document}


\title{\vspace{-2.5cm} Valuation of Variable Annuities with Guaranteed Minimum Withdrawal Benefit under Stochastic Interest Rate}

\author{\vspace{-0.5cm} Pavel V.~Shevchenko$^{1,\ast}$ and Xiaolin Luo$^{2}$}

\date{\vspace{-0.2cm}\footnotesize{Draft, 15 January 2017 (1st version 16 February 2016) }}

\maketitle

\begin{center}
\vspace{-0.65cm}
\footnotesize {\textit{$^{1}$ Applied Finance and Actuarial Studies, Macquarie University, Australia; e-mail: pavel.shevchenko@mq.edu.au  \\
$^{2}$ CSIRO Australia; e-mail: Xiaolin.Luo@csiro.au \\
$^*$ Corresponding author
} }
\end{center}

\begin{abstract}{{
\noindent
This paper develops an efficient direct integration method for pricing of the variable annuity (VA) with guarantees in the case of stochastic interest rate.  In particular, we focus on pricing VA with Guaranteed Minimum
Withdrawal Benefit (GMWB) that promises to return the entire initial
investment through withdrawals and the remaining account balance at maturity. Under the optimal (dynamic) withdrawal strategy of a policyholder, GMWB
pricing becomes an optimal stochastic control problem that can be solved using backward recursion Bellman equation.
Optimal decision becomes a function of not only the underlying asset but also interest rate.
Presently our method is applied to the Vasicek interest rate model, but it is
  applicable to any model when transition density of the underlying asset and interest rate is known in
closed-form or can be evaluated efficiently. Using bond price as a num\'{e}raire the required
expectations in the backward recursion are reduced to
two-dimensional integrals calculated through a high order
Gauss-Hermite quadrature applied on a two-dimensional cubic spline
interpolation. The quadrature is applied after a rotational
transformation to the variables corresponding to the principal axes
of the bivariate transition density, which empirically was observed
to be more accurate than the use of Cholesky transformation.
Numerical comparison demonstrates that the
new algorithm is significantly faster than the partial differential equation or Monte Carlo methods. For pricing of GMWB with dynamic withdrawal strategy, we found that for positive correlation between the underlying
asset and interest rate, the GMWB price under the
stochastic interest rate is significantly higher compared to the
case of deterministic interest rate, while for negative correlation
the difference is less but still significant. In the case of
GMWB with predefined (static) withdrawal strategy, for negative correlation, the difference in prices between
stochastic and deterministic interest rate cases is not material
while for positive correlation the difference is still significant.
The algorithm can be easily adapted to solve similar
stochastic control problems with two stochastic variables possibly
affected by control. Application to numerical pricing of Asian, barrier and
other financial derivatives with a single risky asset under stochastic interest rate is also
straightforward.

 \vspace{0.25cm}
\noindent \textbf{Keywords:} {\emph{Variable annuity, living and death benefits, stochastic interest
rate, optimal stochastic control, Guaranteed Minimum Withdrawal
Benefit, Gauss-Hermite quadrature.}}
}}
\end{abstract}

\pagebreak

\section{Introduction}
\label{sec:introduction} The world population is getting older
fast with life expectancy raising to above 90 years in some countries.
Longevity risk (the risk of outliving one's savings) became critical for retirees.
 Variable annuity (VA) with living and death benefit guarantees is one the
 products that can help to manage this risk. It takes advantage of market growth and at the same time
 provides protection of the savings. VA guarantees are typically classified as
 guaranteed minimum withdrawal benefit (GMWB), guaranteed minimum accumulation benefit (GMAB), guaranteed minimum income benefit (GMIB), and guaranteed minimum death benefit (GMDB). A good overview of VA products and the development
of their market can be found in \cite{bauer2008universal}, \cite{Ledlie2008variableannuities} and \cite{Kalberer2009}.
 Insurers started to sell these types of products from the 1990s in United States. Later, these products
became popular in Europe, UK and Japan. The market of VAs is very large, for example, sales of these contracts in United States between  2011 and 2013 averaged about \$160 billion per year according to the LIMRA (Life Insurance and Market Research Association) fact sheets.

 For clarity and simplicity of presentation, in this paper
we consider a VA contract with
a very basic GMWB guarantee that promises to return the
entire initial investment through cash withdrawals during the policy
life plus the remaining account balance at maturity, regardless of
the portfolio performance. Thus even  when the account of the
policyholder falls to zero before maturity, GMWB feature will
continue to provide the guaranteed cashflows. GMWB allows the policyholder to withdraw funds below or at the contractual rate without
penalty and above the contractual rate with some penalty. If the
policyholder behaves passively and makes withdrawals at the contractual rate defined at
the beginning of the contract, then the behavior of the
 policyholder is  called \emph{\textbf{static}}. In this case the paths of the wealth account  can be simulated  and a standard Monte Carlo (MC)
 simulation method can be used for GMWB pricing. On the other hand if the policyholder optimally decides the amount to withdraw
 at each withdrawal date, then   the behavior of the policyholder is  called \emph{\textbf{dynamic}}.
Under the optimal withdrawal strategy, the pricing
of variable annuities with GMWB becomes an optimal stochastic
control problem. This problem cannot be solved by a  standard
simulation-based method such as the well known Least-Squares MC
method introduced in \citet{Longstaff2001}. This is because the
paths of the underlying wealth process are altered by the optimal
cash withdrawals that should be found from the backward in time solution and the
underlying wealth process cannot be simulated forward in time.
However, it should be possible to apply control
randomization methods extending Least-Square MC to handle optimal
stochastic control problems with controlled Markov processes  recently developed in
\citet{Kharroubi2014}; though the accuracy and robustness of this
method for GMWB pricing has not been studied yet.

It is important to note that the fair fee for the VA guarantee obtained under the assumption that the policyholders  behave optimally to maximise the value of the guarantee is an important benchmark because it is a worst case
scenario for the contract writer. That is, under the no-arbitrage assumption, if the guarantee is perfectly hedged then the issuer will receive a guaranteed profit if the policyholder deviates from the optimal strategy. Pricing under any other strategy will lead to smaller fair fee. Of course, the strategy optimal in this sense may not be optimal to the policyholder under his circumstances and preferences. On the other hand, secondary markets for equity linked insurance products are growing and financial third parties can potentially generate guaranteed profit through hedging strategies from VA guarantees which are not priced according to the worst case assumption about the optimal strategies. There are a number of studies considering these aspects and we refer the reader to \cite{ShevchenkoLuoVAreview2016} for discussion of this topic and references therein.

Pricing of VA with a GMWB feature assuming constant interest rate has been considered in many papers over the last decade. For example,
  \citet{milevsky2006financial} developed a variety of methods for
  pricing GMWB products. In their {\it static} withdrawal approach the GMWB product
  is decomposed into a Quanto Asian put option plus a generic term-certain
  annuity. They also considered pricing when the
  policyholder can terminate (surrender) the contract at the optimal time, which
  leads to an optimal stopping problem akin to pricing an American
  put option. \citet{bauer2008universal}
 presents valuation of variable annuities
with multiple guarantees via a multidimensional discretization approach in
 which  the Black-Scholes partial differential equation (PDE) is  transformed to a
one-dimensional heat equation and a quasi-analytic solution is
obtained through a simple piecewise summation  with a  linear
interpolation on a mesh.
\citet{dai2008guaranteed} developed
an efficient finite difference algorithm using the penalty
approximation to solve the singular stochastic control problem for a
continuous time withdrawal model under the optimal withdrawal
strategy and also finite difference algorithm for discrete time withdrawal.
 Their results show that the GMWB values from the
 discrete time model converge fast to those of the continuous time model. \citet{Huang2012} did a rigorous convergence study of this penalty
 method for GMWB, and \citet{Huang2014} deduce various asymptotes for the free boundaries that separate different
  withdrawal regions in the domain of the GMWB pricing model.
  \citet{Forsyth2008}
present an impulse stochastic control formulation for pricing
variable annuities with GMWB under the optimal policyholder
behavior, and develop a numerical scheme for solving the
Hamilton-Jacobi-Bellman variational inequality for the continuous
withdrawal model  as well as for pricing the discrete withdrawal
contracts.

 More recently, \citet{Azimzadeh2014} prove the existence of an optimal bang-bang control
for a Guaranteed Lifelong Withdrawal Benefits (GLWB) contract. In
particular, they find that the holder of a GLWB can maximize the
contract writer's losses by only performing non-withdrawal, withdrawal
at exactly the contract rate or full surrender. This dramatically
reduces the optimal strategy space. However, they also demonstrate
that the related GMWB contract is not convexity preserving, and
hence does not satisfy the bang-bang principle other than in certain
degenerate cases. GMWB pricing under bang-bang strategy was studied in \cite{LuoShevchenko2015surrender}, and \citet{Huang2015} have developed a regression-based MC
method for pricing GLWB. For GMWB under the optimal withdrawal
strategy, the numerical evaluations have been developed by
\citet{dai2008guaranteed} and \citet{Forsyth2008} using finite difference PDE methods and by \citet{LuoShevchenko2015gmwb} using direct integration method. Pricing of VAs with both GMWB and death benefit (both under static and dynamic regimes) has been developed in \cite{LuoShevchenkoGMWDB2015}.

Some withdrawals from the VA type contracts can also attract country specific government  additional tax and penalty. Recently, \cite{bauer2015behavior} demonstrated that including taxes significantly affects the value of withdrawal guarantees in variable annuities producing results in line with empirical market prices. These matters are not considered in our paper but can be handled by the numerical methodology developed here.

In the literature on pricing GWMB, interest rate is typically assumed to be constant. Few papers considered the case of stochastic interest rate. In particular, \citet{Kwok2012StochInterestGMWB} considered pricing GMWB under the Vasicek stochastic interest rate in the case of \emph{static} withdrawal strategy; they derived the lower and upper bounds for the price because closed-form solution is not available due to withdrawals from the underlying wealth account during its stochastic evolution. \citet{bacinello2011unifying} considered stochastic interest rate and stochastic volatility models under the Cox-Ingersoll-Ross (CIR) models. They developed pricing in the case of static policyholder behavior via the ordinary MC method and  {\it mixed}
  valuation (where the policyholder is \emph{semiactive} and can decide to  surrender the
 contract at any time before the maturity) is
 performed by the Least-Squares MC.
\cite{ForsythVetzal2014} considered modelling stochasticity in the interest rate and volatility via the Markov regime switching models and developed pricing under the static and dynamic withdrawal strategies. Under this approach, the interest rate and volatility are assumed to have the finite
number of possible values and their evolution in time is driven by the finite state Markov chain variable representing possible regimes of the economy.

In this paper, we develop direct integration method for pricing of VAs with guarantees under the
\emph{dynamic} and \emph{static} withdrawal strategies when the interest rate follows
the \emph{Vasicek stochastic interest rate} model. In the case of general stochastic
processes for the underlying asset and interest rate, numerical
pricing can be accomplished by PDE methods that become slow and
difficult to implement in the case of two and more underlying
stochastic variables. Our method is developed for the case when
the bivariate transition density  of the underlying asset and interest
rate are known in closed-form or can be evaluated efficiently. That is, it should be possible to apply this method to the case of, for example, CIR stochastic interest rate model. Using
change of num\'{e}raire technique with bond price as a
num\'{e}raire, the required expectations in the backward recursion
of the stochastic control solution are reduced to the
two-dimensional integrals calculated through a high order
Gauss-Hermite quadrature applied on a two-dimensional cubic spline
interpolation. The quadrature is applied after rotational
transformation to the variables corresponding to the principal axes
of the bivariate transition density which appeared to be more
efficient than the use of the standard Cholesky transformation to
the independent variables.
 For convenience, hereafter  we refer this new algorithm as GHQC (Gauss-Hermite quadrature on cubic spline).
 This allows us to get very fast and accurate results for prices of a typical GMWB contract on the standard desktop computer.
 Previously, in a similar spirit, we developed algorithm for the case of one underlying stochastic risky asset and non-stochastic interest rate  for pricing exotic options in \citet{LuoShevchenkoGHQC2014} and optimal stochastic control problems for pricing GMWB in \citet{LuoShevchenko2015gmwb}.

For clarity of presentation, we focus on pricing of a VA with a very basic GMWB structure. However, the developed methodology can be easily applied to pricing other VA guarantees, see \cite{ShevchenkoLuoVAreview2016} for general formulation of these contracts as the optimal stochastic control problem.
 Finally we would like to mention that the presented algorithm can be easily adapted to solve similar
 stochastic control problems with two state variables possibly affected by control. Also, applications
 to pricing Asian, barrier and other financial derivatives with a single underlying asset under stochastic interest rate are  straightforward.

In the next section we present the underlying stochastic model and describe the GMWB contract.
 Solution as an optimal stochastic control is presented in Section \ref{GMWBasStochControl_sec}.
 Section \ref{PDE_sec} gives a short description of the well known PDE approaches that can be used for pricing.
 Section \ref{GHQC_sec} presents our direct integration GHQC algorithm for
pricing of GMWB contracts under both static and dynamic policyholder behaviors.
 In Section \ref{NumericalResults_sec},  numerical results  for the fair prices and fair fees
 under a series GMWB contract conditions are presented, in comparison with the results from the finite difference method solving corresponding two-dimensional PDEs. The comparison   demonstrates that the new
algorithm produces  results very close to those of the finite
difference PDE method,
  but at the same time  it is significantly faster. Also, the results demonstrate that stochastic interest
  rate has significant impact on price analysed in Section \ref{NumericalResults_sec}. Concluding remarks are given in Section
  \ref{conclusion_sec}. Useful closed-form formulas for the required transition densities, bond and vanilla prices are derived in Appendix \ref{appendixA}.

\section{Model}\label{model_sec}
Following the existing literature, we assume no-arbitrage market with respect to the financial risk and thus the price of the VA with GMWB can be expressed
as an expectation with respect to the risk-neutral probability measure for the underlying
risky asset. Also, there is no mortality risk -- in the event of policyholder death, the contract is maintained by the beneficiary. Death benefit feature commonly offered to the policyholders in addition to GMWB can be easily included into pricing methodology as described in \cite{LuoShevchenkoGMWDB2015}.

Let $(\Omega,\mathcal{F},\mathbb{Q})$ be a probability space with sample space $\Omega$, filtration  $\mathcal{F}=\{\mathcal{F}_t : t\ge 0\}$ (sequence of $\sigma$-algebras $\mathcal{F}_t$ increasing with time $t$ on $\Omega$) and risk-neutral probability measure $\mathbb{Q}$ such that all discounted asset price processes are $\mathbb{Q}$-martingales, i.e. payment streams can be valuated as expected discounted values. Existence of measure $\mathbb{Q}$ implies that the financial market is arbitrage-free and uniqueness of such measure implies that the market is complete. This means that the cost of a portfolio replicating VA contract with guarantee is given by its
expected discounted value under $\mathbb{Q}$. This is a typical set up for pricing of financial derivatives, for a good textbook in this area we refer the reader to e.g. \cite{Bjork2009}.

Consider the joint dynamics for the reference portfolio of assets $S(t)$, e.g. a mutual fund, underlying the contract and the stochastic interest rate $r(t)$, under the risk-neutral probability measure $\mathbb{Q}$, governed by
\begin{eqnarray}\label{referenceportfolio_eq}
\begin{split}
\frac{dS(t)}{S(t)}&=r(t) dt +\sigma_S \left(\rho d\mathcal{B}_1(t) +\sqrt{1-\rho^2} d\mathcal{B}_2(t)\right),\\
dr(t)&=\kappa(\theta-r(t))dt+\sigma_r d\mathcal{B}_1(t).
\end{split}
\end{eqnarray}
Here, $\mathcal{B}_1(t)$ and $\mathcal{B}_2(t)$ are independent standard Wiener processes, $\rho$ is the correlation coefficient between $S(t)$ and $r(t)$ processes, and $\sigma_S$ is the asset volatility parameter. The process for the interest rate $r(t)$ is the well known Vasicek model with constant parameters $\kappa$, $\theta$ and $\sigma_r$. For simplicity of notation we assume that model parameters are constant in time though the results can be generalized to the case of time dependent parameters. We consider time discretization $0=t_0<t_1<\cdots<t_N=T$ corresponding to the contract withdrawal dates, where $t_0=0$ is today and $T$ is the contract maturity.

For this stochastic interest rate model, the price of a zero coupon bond $P(t,T)$ at time $t$ with maturity $T$, can be found in closed-form
\begin{eqnarray}\label{bond_price_eq}
P(t,T)&:=&\mathrm{E}^{\mathbb{Q}}_t\left[e^{-\int_{t}^{T}r(u)du} \right]=e^{A_{t,T}-r(t) B_{t,T}},
\end{eqnarray}
\begin{equation}
B_{t,T}=\frac{1}{\kappa}\left(1-e^{-\kappa (T-t)}  \right),\quad
A_{t,T}=\left(\theta-\frac{\sigma_r^2}{2\kappa^2}\right)\left(B_{t,T}+t-T\right)-\frac{\sigma_r^2}{4\kappa}B_{t,T}^2,\nonumber
\end{equation}
where $\mathrm{E}_{t}^{\mathbb{Q}}[\cdot]$ denotes expectation with respect to the probability measure  $\mathbb{Q}$ conditional on the information available at time $t$. Corresponding stochastic dynamics is easily obtained from (\ref{bond_price_eq}) using It\^{o}'s calculus  to be
\begin{equation}\label{bond_sde}
\frac{dP(t,T)}{P(t,T)}=r(t) dt -\sigma_r B_{t,T} d\mathcal{B}_1(t).
\end{equation}
Solution for the process (\ref{referenceportfolio_eq}), which is a bivariate Normal distribution for $(\ln S(t), r(t))$ given $(S(0), r(0))$, and the bond price formula are derived in Appendix \ref{appendixA}.


Consider the following VA contract with a basic GMWB often used in research studies which is convenient for benchmarking. The actual products may have extra features but these can be easily incorporated in the model and numerical algorithm developed in this paper.
\begin{itemize}
\item The premium paid by the policyholder upfront at $t_0$ is invested into the reference portfolio/risky asset $S(t)$.
The value of this portfolio (hereafter
referred to as \emph{\textbf{wealth account}}) at time $t$ is denoted as $W(t)$, so that
the upfront premium paid by the policyholder is $W(0)$. GMWB guarantees
the return of the premium via the withdrawals $\gamma_n\ge 0$ allowed at
times $t_n$, $n=1,2,\ldots,N$.
Let $N_w$ denote the number of
withdrawals per annum.
The total of withdrawals cannot exceed the guarantee $W(0)$
and withdrawals can be different from the contractual (guaranteed)
withdrawal $G_n=W(0)(t_n-t_{n-1})/T$,  with penalties imposed if
$\gamma_n>G_n$. Denote the annual contractual rate as $g:=1/T$.
Then the wealth account $W(t)$ evolves as
\begin{eqnarray}\label{eq_Wt}
\begin{split}
W(t_n^-)&=\frac{W(t_{n-1}^+)}{S(t_{n-1})}S(t_n) e^{-\alpha \Delta_n},\\
W(t_n^+)&=\max\left(W(t_n^-)-\gamma_n,0\right),\;\; n=1,2,\ldots,N,
\end{split}
\end{eqnarray}

where $\Delta_n=t_n-t_{n-1}$ and \emph{$\alpha$ is the annual fee continuously charged by the contract issuer}. If the account balance
becomes zero or negative, then it will stay zero till maturity.
The process for $W(t)$ within $(t_{n-1},t_n)$ is the same as the process for the
underlying asset $S(t)$ in (\ref{referenceportfolio_eq}) except that the drift term $r(t)$ is replaced by $r(t)-\alpha$.

\item Denote the value of the contract guarantee at time $t$ as $A(t)$, hereafter referred to as \emph{\textbf{guarantee account}}, with $A(0)=W(0)$. Hereafter, denote the time immediately before $t_n$ (i.e. before withdrawal) as  $t_n^-$, and  immediately
after $t_n$ (i.e. after withdrawal) as  $t_n^+$ and let all functions discontinuous at $t_n$ be right-continuous with finite left limit.
The guarantee balance
evolves as
\begin{equation}\label{accountbalance_eq}
A(t_n^+)=A(t_n^{-})-\gamma_n=A(t^+_{n-1})-\gamma_n,\;\; n=1,2,\ldots,N
\end{equation}
with $A(T^+)=0$, i.e. $W(0)=A(0) \ge \gamma_1+\cdots+\gamma_N$ and
$A(t_{n-1}^{+})\ge \sum_{k=n}^N\gamma_{k}$. The account balance $A(t)$ remains
unchanged within the interval $(t_{n-1},\;t_n), \;n=1,2,\ldots,N$.

\item The cashflow received by the policyholder at the withdrawal time $t_n$ is given by
\begin{equation}
C_n(\gamma_n)=\left\{\begin{array}{ll}
                   \gamma_n, & \mathrm{if}\; 0\le \gamma_n\le G_n, \\
                   G_n+(1-\beta)(\gamma_n-G_n), & \mathrm{if}\; \gamma_n>G_n,
                 \end{array} \right.
\end{equation}
where $G_n$ is the contractual  withdrawal and $\beta\in
[0,1]$ is the penalty coefficient applied to the portion of withdrawal above
$G_n$.

\item Let $Q_t(W,r,A)$ be a price of the  VA contract with GMWB at time $t$, when $W(t)=W$, $r(t)=r$, $A(t)=A$.  At maturity, the policyholder takes the maximum between
the remaining guarantee account net
of penalty charge and the remaining balance of the wealth account,
i.e. the final payoff is
\begin{equation}\label{finalcond_eq}
Q_{t_N^-}(W,r, A)=\max\left(W,C_N(A)\right).
\end{equation}

\end{itemize}

During the contract, the policyholder receives cashflows $C_n(\gamma_n)$,
$n=1,2,\ldots,N-1$ and the final payoff at maturity. Denote the Markov state vector at time $t$ as $V_t=(W(t),r(t),A(t))$ and $\bm{V}=(V_t)_{0\le t\le T}$.
Given the withdrawal strategy $\bm{\gamma}=(\gamma_1,\ldots,\gamma_{N-1})$, the present
value of the total contract payoff is
\begin{equation}
H_0(\bm{V},\bm{\gamma})=e^{-\int_{0}^{T} r(\tau)d\tau} \max\left(W(T^-),C_N(A(T^-))\right)+\sum_{n=1}^{N-1}
e^{-\int_{0}^{t_n} r(\tau)d\tau}C_n(\gamma_n).
\end{equation}
Under the above assumptions/conditions, the fair no-arbitrage value of the contract for the pre-defined (\emph{\textbf{static}}) withdrawal strategy $\bm{\gamma}$ can be calculated as
\begin{equation}\label{GMWBstatic_eq}
Q_0\left(V_0\right)=\mathrm{E}_{t_0}^{\mathbb{Q}}\left[H_0(\bm{V},\bm{\gamma})\right].
\end{equation}

Under the optimal (\emph{\textbf{dynamic}}) withdrawal strategy, where the decision on withdrawal amount $\gamma_n$ is based upon the information $\mathcal{F}_{t_n}$ available at time $t_n$, the fair contract value is
\begin{equation}\label{GMWB_general_eq}
Q_0\left(V_0\right)=\sup_{\bm{\gamma}}\mathrm{E}_{t_0}^{\mathbb{Q}}\left[H_0(\bm{V},\bm{\gamma})\right],
\end{equation}
where $\gamma_{1},\ldots,\gamma_{N-1}$ are the control variables (withdrawals) chosen to maximize the expected value of discounted cashflows and supremum is taken over all admissible strategies. Note that the withdrawal $\gamma_n:=\gamma_n(V_{t_n^-})$ at time $t_n$ is a function of the state variable $V_{t_n^-}$, i.e. it can be different for different realizations of $V_{t_n^-}$.
Moreover the control variable $\gamma_n$ affects the transition law of the underlying wealth process from $t_n^-$ to $t_{n+1}^-$. Any strategy different from optimal is \emph{sub-optimal} and leads to a smaller price.

The today's value of the contract $Q_0(V_0)$ is a function of a fee $\alpha$ charged by the issuer for GMWB guarantee. The fair fee value of $\alpha$ to be charged for providing GMWB feature corresponds to $Q_0(V_0)=W(0)$. That is, once a pricing of $Q_0(V_0)$ for a given value of $\alpha$ is developed, then a numerical root search algorithm is required to find the fair fee.

It is important to note that the fair fee for the VA guarantee obtained under the assumption that the policyholders  behave optimally to maximise the value of the guarantee is a \emph{worst case scenario for the contract writer}. If the guarantee is perfectly hedged then the issuer will receive a guaranteed profit if the policyholder deviates from the optimal strategy. Pricing under any other strategy will lead to smaller fair fee. Of course, the strategy optimal in this sense may not be optimal to the policyholder under his circumstances and preferences but it is an important benchmark.

In practice, there will be a residual risk due to discrete in time hedging and incompletenesses of financial market that can be handled by adding extra loading on the price under the actuarial approach or adjusting risk premium under the no-arbitrage financial mathematics approach so that the risk of hedging error loss will not exceed the required level. These adjustments depend on the risk management strategy for the product and will not be considered here; for discussion and references, see e.g. \cite{ShevchenkoLuoVAreview2016}.

\begin{remark}
Note that we started our modelling with assumption of the stochastic model (\ref{referenceportfolio_eq}) under the risk-neutral probability measure $\mathbb{Q}$. For risk management purposes, one might be interested to start with the process under the real (physical) probability measure $\mathbb{P}$,
\begin{eqnarray*}
\begin{split}
\frac{dS(t)}{S(t)}&=\mu^\ast(t) dt +\sigma_S \left(\rho d\mathcal{B}^\ast_1(t) +\sqrt{1-\rho^2} d\mathcal{B}^\ast_2(t)\right),\\
dr(t)&=u^\ast(r,t)dt+\sigma_r d\mathcal{B}^\ast_1(t),
\end{split}
\end{eqnarray*}
with independent standard Wiener processes $\mathcal{B}^\ast_1(t)$ and $\mathcal{B}^\ast_2(t)$, and derive corresponding
risk-neutral process (\ref{referenceportfolio_eq}) for the VA guarantee and bond price valuation. This can be done in the usual way by forming a portfolio $\Pi_t=-U_t(W,r,A)+\Delta_S\times S+\Delta_P\times P(t,T)$, where $U_t(W,r,A):=Q_t(W,r,A)-W$ is the value of VA guarantee, $\Delta_S$ is the number of units of $S(t)$ and $\Delta_P$ is the number of units of bond $P(t,T)$. Then calculate the change of portfolio $d\Pi_t$ using Ito's lemma, and set
$$\Delta_S=\frac{W}{S}\frac{\partial U_t(W,r,A)}{\partial W}\quad \mbox{and}\quad \Delta_P=\frac{\partial U_t(W,r,A)/\partial r}{\partial P(t,T)/\partial r}$$
to eliminate random terms so that the portfolio earns risk free interest rate $d\Pi_t=r\Pi_t dt$. This leads to a PDE (\ref{eqn_pde}) for $Q_t(W,r,A)$ and using Feynman-Kac theorem one can establish that the process corresponding to this PDE  is the risk-neutral process (\ref{referenceportfolio_eq}). For details, see e.g. \cite[section 6.5]{ShevchenkoLuoVAreview2016} and textbook \cite[sections 30.3 and 33.6]{wilmott2006paul}. It is important to note that this procedure will also introduce the market price of interest rate risk $\lambda(r,t)$ such that the drift of the interest rate risk-neutral process is $u^\ast(r,t)-\lambda(r,t)\sigma_r$; then under the assumption that $\lambda(r,t)$ and $u^\ast(r,t)$ are linear functions of $r$ one can write the risk-neutral process for $r$ as in (\ref{referenceportfolio_eq}).
\end{remark}

\section{Pricing GMWB as optimal stochastic control}\label{GMWBasStochControl_sec}
Given that the discrete in time state vector $V_{t_n^-}=(W(t_n^-),r(t_n^-),A(t_n^-))$, $n=0,1,\ldots,N$ is a Markov process, it is easy to recognize that the contract valuation under the optimal withdrawal strategy (\ref{GMWB_general_eq}) is the optimal stochastic control problem for controlled Markov process that can be solved  recursively to find the contract value $Q_{t_n^-}(\cdot)$ at $t_n^-$, $n=N-1,\ldots,0$ via the well known backward induction Bellman equation
\begin{eqnarray}\label{Bellman_eq}
&&Q_{t_n^-}(W(t_n^-),r(t_n),A(t_n^-))=\sup_{0\le\gamma_n\le A(t_{n}^-)}\bigg(C_n(\gamma_n)\nonumber\\
&&\;\;\;\quad+\mathrm{E}^{\mathbb{Q}}_{t_{n}^+}\left[e^{-\int_{t_n}^{t_{n+1}}r(\tau)d\tau}Q_{t_{n+1}^{-}}\left(W(t_{n+1}^-), r(t_{n+1}), A(t_{n+1}^-)\right)\bigg|W(t_{n}^+),r(t_n), A(t_n^+)\right]\bigg)
\end{eqnarray}
starting from the final condition $Q_{t_N^-}(W,r,A)=\max\left(W,C_N(A)\right)$.
For a good textbook treatment of stochastic control problem in finance, see \cite{bauerle2011markov}.
Static pricing (\ref{GMWBstatic_eq}) under the predefined strategy $\bm\gamma$ can be also done using the above backward induction with supremum removed.


 For each $t_n$, $n=1,\ldots,N-1$, this backward recursion (\ref{Bellman_eq}) involves calculation of the expectation
\begin{eqnarray}\label{eq_expS}
Q_{t^+_{n}}\left(W,  r, A\right)=\mathrm{E}^{\mathbb{Q}}_{t_{n}^+}\left[e^{-\int_{t_{n}}^{t_n+1} r(\tau) d\tau}
Q_{t_{n+1}^{-}}\left(W(t_{n+1}^-),r(t_{n+1}), A(t_{n+1}^-)\right)|W, r, A\right]
\end{eqnarray}
and application of the jump condition across $t_{n}$
\begin{equation}\label{eqn_jump}
Q_{t_{n}^-}(W,r, A)=\max_{0 \leq \gamma_n\leq A } [C_n(\gamma_n)+Q_{t_n^+}(\max(W-\gamma_n,0), r,A-\gamma_n)].
\end{equation}

Calculating expectation (\ref{eq_expS}) is difficult as it would require three-dimensional integration with respect to the joint distribution of three random variables $W(t_{n+1}^-)$, $r(t_{n+1})$ and $Y(t_{n+1})=\int_{t_{n}}^{t_{n+1}}r(u)du$ conditional on $W({t_{n}^+})$ and $r({t_{n}})$; note that variable $A(t)$ does not change within $(t_{n},t_{n+1})$. Actually the required 3$d$ distribution can be found in closed-form in the case of stochastic process (\ref{referenceportfolio_eq}) considered here, see Appendix \ref{appendixA}, which is useful for validation of calculations in the case of static withdrawals via direct simulation of process (\ref{referenceportfolio_eq}).
 However if we change num\'{e}raire from the money market account $M(t)=e^{\int_0^t r(\tau)d\tau}$ to  the bond $P(t_{n},t_{n+1})$ with maturity $t_{n+1}$, i.e. change probability measure with Radon-Nikodym derivative
 \begin{equation}
 \mathbb{Z}_t=\left.\frac{d\widetilde{\mathbb{Q}}}{d\mathbb{Q}}\right|_{\mathcal{F}_t}=\frac{M(t_{n})}{M(t)}\frac{P(t,t_{n+1})}{P(t_{n},t_{n+1})},\quad t\in [t_{n},t_{n+1}],
 \end{equation}
 then the expectation (\ref{eq_expS}) simplifies to the two-dimensional integration
 \begin{eqnarray}\label{optionprice_new}
 \begin{split}
& \mathrm{E}^{\mathbb{Q}}_{t_{n}^+}\left[e^{-\int_{t_{n}}^{t_{n+1}}r(u)du}Q_{t_{n+1}^-}\left(W(t_{n+1}^-),r(t_{n+1}), \cdot\right)\bigg|\;\cdot\;\right]\\
 &\quad\quad \quad\quad           =P(t_{n},t_{n+1}){\mathrm{E}}^{\widetilde{\mathbb{Q}}}_{t_{n}^+}\left[Q_{t_{n+1}^-}\left(W(t_{n+1}^-),r(t_{n+1}), \cdot\right)\bigg|\;\cdot\;\right],
 \end{split}
 \end{eqnarray}
 where ${\mathrm{E}}^{\widetilde{\mathbb{Q}}}_{t_{n}}[\cdot]$ is expectation under the new probability measure $\widetilde{\mathbb{Q}}$. The process for $\mathbb{Z}_t$ is easily obtained from the process (\ref{bond_sde}) for the bond price $P(t,t_{n+1})$ as
 $$
 d\mathbb{Z}_t=\phi(t)\mathbb{Z}_t d\mathcal{B}_1,\quad \phi(t)=-\sigma_r B_{t,t_{n+1}}.
 $$
  Then, using Girsanov theorem the required transformation to the Wiener process is $\mathcal{B}_1(t)=\phi(t)dt+d\widetilde{\mathcal{B}}_1(t)$, and the processes under the new measure $\widetilde{\mathbb{Q}}$ for $t\in (t_{n},t_{n+1})$ are
 \begin{equation}\label{new_riskneutralprocess}
 \begin{split}
dS(t)/S(t)&=(r(t)+\sigma_S \rho \phi(t)) dt +\sigma_S \left(\rho d\widetilde{\mathcal{B}}_1(t) +\sqrt{1-\rho^2} d\widetilde{\mathcal{B}}_2(t)\right),\\
dr(t)&=\kappa\left(\widetilde\theta(t)-r(t)\right)dt+\sigma_r d\widetilde{\mathcal{B}}_1(t);\;\widetilde\theta(t)=\theta+\frac{\sigma_r}{\kappa}\phi(t)
\end{split}
\end{equation}
with $\widetilde{\mathcal{B}}_1(t)$ and $\widetilde{\mathcal{B}}_2(t)$ independent Wiener processes.  Note that $\phi(t)$ is volatility of the bond $P(t,t_{n+1})$, see (\ref{bond_sde}). Solution for this process, which is a bivariate Normal distribution for $(\ln S(t), r(t))$ given $(S(0), r(0))$, is derived in Appendix \ref{appendixA}. For a good textbook treatment of change of num\'{e}raire technique, see \cite[chapter 26]{Bjork2009}.
It is important to note that for different time steps, the change of measure is based on bonds of different maturities.


Assuming the probability
density function of $W(t_{n+1}^-)$ and $r(t_{n+1})$ at $W(t_{n+1}^-)=w^\prime$ and $r(t_{n+1})=r^\prime$ conditional on $W(t_{n}^+)=w$ and $r(t_{n})=r$ under the new probability measure $\widetilde{\mathbb{Q}}$  is known in closed-form
$p_{n+1}(w^\prime,r^\prime|w,r)$, the required expectation (\ref{optionprice_new}) can be evaluated
as
\begin{equation}\label{eq_intS}
Q_{t_{n}^+}\left(w, r, A\right)=P(t_{n},t_{n+1})\int\int
p_{n+1}\left(w^\prime,r^\prime|w,r\right) Q_{t_{n+1}^-}(w^\prime,r^\prime,A)dw^\prime dr^\prime.
\end{equation}
In the case of underlying stochastic process (\ref{referenceportfolio_eq}) the transition density $p_{n+1}(w^\prime,r^\prime|w,r)$ is known in closed-form and we will use the Gauss-Hermite quadrature for  evaluation of the
above integration over an infinite domain. The required continuous
function $Q_t(W,r, A)$ will be approximated by a two-dimensional cubic spline
interpolation on a discretized grid in the  $(W,r)$ space. Note, in general, a three-dimensional
interpolation in the $(W,r, A)$ space is required, but one can manage to avoid interpolation in $A$ by ``smart" numerical manipulation setting the jump amounts in $A$ spaced in such a way that the $A$ reduced after jump is always on a grid point.
Below we discuss details of the algorithm of  the numerical
integration of (\ref{eq_intS}) using  Gauss-Hermite quadrature on a
cubic spline interpolation, followed by the application of jump
condition (\ref{eqn_jump}).

%
 \emph{Note, for simple options/contracts where payoff depends on the underlying asset only and is received at the contract maturity, a change of num\'{e}raire can remove stochastic interest rate dimension from pricing, effectively reducing numerical problem to
the deterministic interest rate case. However, for pricing GMWB either static or dynamic cases, the
additional dimension in the interest rate $r$ cannot be avoided due to withdrawals (jump conditions) during the contract life.}

\section{Numerical valuation of GMWB via PDE}\label{PDE_sec}
In the case of continuous in time withdrawal, following the procedure of
deriving the Hamilton-Jacobi-Bellman (HJB) equations in stochastic
control problems, the value of the VA contract with guarantee under the optimal withdrawal
is found to be governed by a two-dimensional PDE in the case of deterministic interest rate; see \citet{milevsky2006financial},
 \citet{dai2008guaranteed} and \citet{Forsyth2008}, that will become
 three-dimensional PDE in the case of stochastic interest rate. For
discrete withdrawals, the governing PDE in the period between
withdrawal dates is one dimension less than the continuous case because the guarantee
account balance $A(t)$ remains unchanged between withdrawals, similar to the Black-Scholes
equation, with jump conditions at each withdrawal date to link the
prices at the adjacent periods.
In particular, the contract value
$Q_t(W,r,A)$ at $t\in (t_{n-1},t_n)$ satisfies
\begin{equation}\label{eqn_pde}
\frac{\partial Q_t}{\partial
t}+\frac{\sigma_S^2}{2}W^2\frac{\partial^2 Q_t}{\partial
W^2}+(r-\alpha)W\frac{\partial Q_t}{\partial W}+\frac{\sigma_r^2}{2}\frac{\partial^2 Q_t}{\partial
r^2}+(\theta-r)\frac{\partial Q_t}{\partial r}+\rho\sigma_S\sigma_r W\frac{\partial^2 Q_t}{\partial W\partial r} -rQ_t=0,
\end{equation}
 that can be solved numerically using e.g. Crank-Nicholson finite difference scheme for
 each $A$ backward in time with jump condition (\ref{eqn_jump}) applied at withdrawal dates $t_n$.
  A more efficient and very popular class of algorithms is the \emph{alternating direction implicit}
  (ADI) method, among which a standout variation is the so called \emph{hopscotch method},
  introduced by \citet{Gourlay1970} by a reformulation of an idea of \citet{Gordon1965}.
  It was shown that  hopscotch method was an ADI process with  a novel way of decomposing the problem
  into simpler parts. The general idea is to solve alternative points explicitly and then employ
  an implicit scheme to solve for the remaining points explicitly. The original  hopscotch method
  cannot be applied readily to equations with mixed derivatives without introducing a certain amount
   of implicitness. \citet{Gourlay1977} suggested two techniques for dealing with the mixed
   derivative -- \emph{ordered odd-even hopscotch} and \emph{line hopscotch}. Numerical tests in
    \citet{Gourlay1977} showed that the line hopscotch performed best for both constant and
    variable coefficient parabolic equation cases, in comparison with the ordered odd-even hopscotch
    and a locally one dimensional (LOD) methods.

    In this work, for numerical validation of our GHQC algorithm, we have implemented the line hopscotch method. In brief,
 assuming the parabolic equation is discretized with the finite difference grid points $(x_i, y_j,t_n)$,
 the line hopscotch method first explicitly evaluates the solution at those points which have $(n+j)$ even,
 and then solves implicitly for those points with $(n+j)$ odd. The alternative value of $j$ for a given
  time step $n$ gives a tri-diagonal set of equations, provided the finite difference operators are
  chosen in a certain manner. For details, see \citet{Gourlay1977}.

\section{GHQC direct integration method}\label{GHQC_sec}
In this section we present details of the algorithm for numerical
integration (\ref{eq_intS}) using  the Gauss-Hermite quadrature on a
cubic spline interpolation, followed by the application of jump
condition (\ref{eqn_jump}), referred to as GHQC.

\subsection{Algorithm structure}
Our approach  relies on computing expectations (\ref{eq_intS}) in a backward time-stepping between
 withdrawal dates through a high order Gauss-Hermite integration quadrature applied on a cubic spline
interpolation. It is easier to implement and computationally faster than PDE method in the case of transition density of underlying stochastic variables known in closed-form.
For a given guarantee account variable $A$ within $(t_n,t_{n+1})$,
the price $Q_{t_n^+}(W,r,A)$  can be numerically evaluated  using
(\ref{eq_intS}).  For now we leave details of computing
(\ref{eq_intS}) to the next section and assume it can be done with
sufficient accuracy and efficiency. Starting from a final condition
at $t=t_N^-$ (just immediately before the final withdrawal), a
backward time stepping using (\ref{eq_intS}) gives solution at
$t=t_{N-1}^+$. Applying jump condition (\ref{eqn_jump}) to the
solution at $t=t_{N-1}^+$ we obtain the solution at $t=t_{N-1}^-$
from which further backward time stepping gives us solution at
$t_0$ to find $Q_0(W(0),r(0),W(0))$.  In order to
 apply the jump condition at each withdrawal date,
 the solution has to be found for many different
levels of $A$.  The numerical algorithm takes the
following key steps.

\begin{itemize}
\item Step 1. Generate an auxiliary finite grid  $0 = A_1 < A_2 <
\cdots < A_J = W(0)$ to track solutions for different values of the guarantee  account $A$. Discretize the wealth account $W$ space as $W_0
,W_1, \ldots,W_M$ and the interest rate $r$ space as $r_0
,r_1, \ldots,r_K$.
\item Step 2. At $t=t_N^-$, initialize $Q_{t_N^-}(W, r, A)$ with a given continuous payoff function at maturity (\ref{finalcond_eq})
  required by the following step of integration.
\item Step 3. For $t=t^+_{N-1}$, evaluate integration (\ref{eq_intS}) for each node point $(W_m, r_k, A_j)$ and using a
one-dimensional cubic spline interpolation
 in $W$ to obtain the continuous function $Q_{t_{N-1}^+}(W, r_k, A_j)$ required by the following step of applying the jump condition.
\item Step 4. Apply the jump condition (\ref{eqn_jump}) for all possible withdrawals $\gamma_{N-1}$
and find the withdrawal maximizing $Q_{t_{N-1}^-}(W_m, r_k, A_j)$
for all grid points $j=1,\ldots, J$, $k=0,\ldots, K$
 and $m=0,\ldots, M$. Use two-dimensional cubic spline
 interpolation to obtain continuous function $Q_{t_{N-1}^-}(W, r,
 A_j)$ required by  the next step of integration.
\item Step 5. Repeat Step 3 and Step 4 for $t=t_{N-2}, t_{N-3}, \ldots, t_1$.
\item Step 6. Evaluate integration (\ref{eq_intS}) for the backward time step from $t_1^-$ to $t_0$ for the single
point $(W(0), r(0), A(0))$ to obtain solution $Q_0(W(0), r(0),
A(0))$ for the contract price at $t=t_0$.
\end{itemize}

\subsection{Numerical evaluation of the expectation}\label{sec_GHQC}
Similar to a finite difference scheme, we discretize the
wealth space domain $[W_{\min}, W_{\max}] $  as $W_{\min} =W_0 < W_1<
\cdots<W_M=W_{\max}$ , where $W_{\min}$ and $W_{\max}$ are the lower
and upper boundary respectively. Similarly, the interest rate space is
  discretized as $r_{\min} =r_0 < r_1<
\ldots<r_K=r_{\max}$ , where $r_{\min}$ and $r_{\max}$ are the bounds for the interest rate.

 For pricing GMWB, due to the
finite reduction of $W$ at each withdrawal date, we have to consider
the possibility of zero $W$, thus the lower bound
$W_{\min}=0$. The upper bound is set  sufficiently
 far from the
initial value at time zero $W(0)$. In general for both $W$ and $r$ dimensions,
the proper choice of the lower and upper bounds is  guided by the joint
distribution of $\ln S(T)$ and $r(T)$, derived in Appendix
\ref{appendixA}, to ensure that the probability for the random
process to go beyond the bounds is immaterial.

 The idea
is to find the contract values at all  grid points at  each time
step
 $(t_{n-1}^+,t_n^-)$  through integration  (\ref{eq_intS}),
starting at maturity $t=t_N^-$.  At each time step we evaluate
the integration (\ref{eq_intS}) for every grid point  by a high
accuracy numerical quadrature.

Under the new probability measure $\widetilde{\mathbb{Q}}$, the process for $\ln W(t)$ and $r(t)$  between the withdrawal dates is a simple Gaussian
 process given by (\ref{eq_Wt}) and (\ref{new_riskneutralprocess}), where the conditional joint density of
$(\ln W(t_n^-), r(t_n))$ given $\ln W(t_{n-1}^+)=x^\ast$, $r(t_{n-1})=r^\ast$  is a bivariate Normal density function,
as shown in Appendix \ref{appendixA}, with the mean, variance and covariance given by
\begin{subequations}\label{mean_cov_eq}
\begin{align}
\mu_r(r^\ast):&=\mathrm{mean}(r({t_{n}}))=r^\ast e^{-\kappa \Delta_n}+\left(\theta-\frac{\sigma_r^2}{\kappa^2}\right)b_n+\frac{\sigma_r^2}{2\kappa^2}a_n;\displaybreak[0]\\
\tau_r^2:&=\mathrm{var}(r({t_{n}}))=\frac{\sigma_r^2}{2\kappa}a_n;\displaybreak[0]\\
\mu_x(x^\ast,r^\ast):&=\mathrm{mean}(\ln W(t_{n}^-))=
x^\ast +\frac{b_n}{\kappa}\left(r^\ast+\frac{b_n\sigma_r^2}{2\kappa^2}\right)+\left(\theta-\frac{\sigma_r^2}{\kappa^2}\right)\left(\Delta_n-\frac{b_n}{\kappa}\right)
\displaybreak[0]\\
&\quad\quad\quad\quad\quad\quad\quad\quad -\frac{\rho \sigma_S \sigma_r}{\kappa^2}(\kappa \Delta_n- b_n)-\left(\alpha+\frac{1}{2}\sigma_S^2\right)\Delta_n;\displaybreak[0]\displaybreak[0]\\
\tau_x^2:&=\mathrm{var}(\ln W(t_{n}^-))=\sigma_S^2
\Delta_n+\frac{\sigma_r^2}{2\kappa^3}(2\kappa\Delta_n-4b_n+a_n)
+\frac{2\rho\sigma_S\sigma_r}{\kappa^2}\left(\kappa \Delta_n-b_n\right);\displaybreak[0]\\
\rho_{xr}:&=\frac{\mathrm{cov}(\ln W(t_{n}^-),r(t_{n}))}{\tau_x\tau_r},\quad
\mathrm{cov}(\ln W(t_{n}^-),r(t_{n}))= \frac{\rho\sigma_S\sigma_r
b_n}{\kappa}+\frac{\sigma_r^2}{2\kappa^2}(2b_n-a_n),
\end{align}
\end{subequations}
where $b_n=1-e^{-\kappa \Delta_n}$, $a_n=1-e^{-2\kappa \Delta_n}$ and $\Delta_n=t_n-t_{n-1}$. For simplicity, here we omit time step index $n$ in notation for the means and covariances.

%
Thus the density of ${Y_1}=(\ln W(t_n^-)-\mu_x)/\tau_x$ and
${Y_2}=(r(t_n)-\mu_r)/\tau_r$ is the standard bivariate Normal with
zero means, unit variances, and correlation $\rho_{xr}$. If we
apply the change of variables
\begin{equation}\label{eq_transZt}
{Z_1}=\frac{{Y_1}}{\sqrt{2(1-\rho_{xr}^2)}},\;\;\;
{Z_2}=\frac{{Y_2}}{\sqrt{2(1-\rho_{xr}^2)}}
\end{equation}
the integration (\ref{eq_intS}) becomes
\begin{equation}\label{eq_intY}
\begin{split}
&Q_{t_{n-1}^+}\left(W, r, A\right)\\
&\quad\quad=P(t_{n-1},t_n)\frac{\sqrt{1-\rho_{xr}^2}} {\pi}
\int_{-\infty}^{+\infty}\int_{-\infty}^{+\infty}
e^{-{z}_1^2-{z}_2^2-2\rho_{xr}{z}_1{z}_2}
Q^{({Z})}_{t_n^-}({z}_1,{z}_2,A) d{z}_1 d{z}_2
\end{split}
\end{equation}
which has the form suitable for integration using the Gauss-Hermite
quadrature. Here, $Q^{({Z})}_t(\cdot)$ denotes $Q_t(\cdot)$ as a function of $Z_1$ and $Z_2$ after transformation from $W$ and $r$.

For an arbitrary one-dimensional function $f(x)$, the Gauss-Hermite
quadrature is applied as
\begin{equation}\label{eq_GHQ}
\int_{-\infty}^{+\infty}e^{-x^2}f(x)dx \approx \sum_{i=1}^q
\lambda_i^{(q)} f(\xi_i^{(q)}),
\end{equation}
where  $q$ is the order of the Hermite polynomial, $\xi_i^{(q)}$, $i = 1,2,\ldots,q$ are
the roots of the Hermite polynomial $H_q(x)$, and
the associated weights $ \lambda_i^{(q)}$  are given by
$$\lambda_i^{(q)}= \frac {2^{q-1} q! \sqrt{\pi}} {q^2\left(H_{q-1}(\xi_i^{(q)})\right)^2}.$$
This approximation is exact when $f(x)$ can be represented as
polynomial of the order up to $2q-1$. In general, the abscissas
$\xi_i^{(q)}$ and the weights $\lambda_i^{(q)}$ for the
Gauss-Hermite quadrature for a given order $q$ can be readily
computed, e.g. using functions in \citet{Pres92}, or available in
precalculated tables.

Decomposing the two-dimensional integration in (\ref{eq_intY}) into nested one-dimensional integration and  applying  the one-dimensional
Gauss-Hermite quadrature to each of the  variable, we obtain
  \begin{equation}\label{eq_qy}
  \begin{split}
&\int_{-\infty}^{+\infty}\int_{-\infty}^{+\infty}
e^{-{z}_1^2-{z}_2^2-2\rho_{xr}{z}_1{z}_2} Q^{({Z})}_{t_n^-}({z}_1,{z}_2,A) d{z}_1 d{z}_2\\
&\quad\quad\quad\quad\approx   \sum_{i=1,j=1}^{q_1,q_2}
\lambda_i^{(q_1)} \lambda_j^{(q_2)}
e^{-2\rho_{xr}\xi_i^{(q_1)}\xi_j^{(q_2)}}
Q_{t_n^-}^{(Z)}(\xi_i^{(q_1)},\xi_j^{(q_2)},A).
\end{split}
\end{equation}
Note, in general,  different orders $q_1$ and $q_2$ can be used. For
example we might let $q_1>q_2$ to take into consideration that the
value function $Q_t^{(Z)}(z_1,z_2,\cdot)$ changes more rapidly with $z_1$ than with $z_2$,
thus making the quadrature points more efficiently assigned to the
variables.

Unfortunately, the numerical integration (\ref{eq_qy}) is not efficient due to presence of the
factor $e^{-2\rho_{xr}z_1 z_2}$, when there is a
non-zero correlation between stock market and the interest rate.
This can be improved by transformation to independent random
variables ($Z_1$,$Z_2$) using the standard Cholesky transformation
\begin{equation}\label{cholseky_eq}
  Y_1=\sqrt{2}Z_1,\quad Y_2=\sqrt{2}(\rho_{xr} Z_1+\sqrt{1-\rho_{xr}^2}Z_2).
\end{equation}
However, we observed that more accurate results are obtained by a
transformation to independent variables ($Z_1$, $Z_2$) corresponding
to the principal axis of the joint density, which can be done using
a matrix spectral decomposition
\begin{equation}\label{rotation_eq}
  Y_1=\sqrt{2}(aZ_1+bZ_2),\quad Y_2=\sqrt{2}(bZ_1+aZ_2),
\end{equation}
where
$$a=\frac{1}{2}(\sqrt{1+\rho_{xr}} + \sqrt{1-\rho_{xr}}),\quad b=\frac{1}{2}(\sqrt{1+\rho_{xr}} - \sqrt{1-\rho_{xr}}).$$
Using transformation (\ref{rotation_eq}), not only the cross term in the density disappears,
but also it is standardized for applying the Gauss-Hermite quadrature. Now,
in terms of the new variables $(Z_1, Z_2)$, the integration (\ref{eq_qy}) changes  to
a simpler but more accurate approximation
 \begin{equation}\label{eq_qz}
 \begin{split}
\int_{-\infty}^{+\infty}\int_{-\infty}^{+\infty}
e^{-z_1^2-z_2^2} Q^{(Z)}_{t_n^-}(z_1,z_2,A) dz_1 dz_2\approx   \sum_{i=1}^{q_1}\sum_{j=1}^{q_2}
\lambda_i^{(q_1)} \lambda_j^{(q_2)}  Q_{t_n^-}^{(Z)}\left(\xi_i^{(q_1)},\xi_j^{(q_2)},A\right).
\end{split}
\end{equation}

If we apply the change of variable  and the
Gauss-Hermite quadrature (\ref{eq_qz}) as described above to every grid point $(W_m, r_k, A_j)$,
$m=0,1,\ldots,M$, $k=0,1,\ldots,K$ and $j=0,1,\ldots,J$, i.e. let $W(t_{n-1}^+)=W_m$,  $r(t_{n-1}^+)=r_k$ and $A=A_j$, then the  contract values
at time $t=t_{n-1}^+$  for all the grid points can be evaluated.

As is commonly practiced, we select the working domain in the asset space to be in
 terms of $X=\ln (W/W(0))$, i.e. we set
 $X_{\min}=\ln(W_{\min}/W(0))$ and $X_{\max}=\ln(W_{\max}/W(0))$. The domain $[X_{\min}, X_{\max}]$ is
 uniformly discretised with step $\delta X=(X_{\max}-X_{\min})/M$ to yield the grid $X_m=X_{\min}+m\delta X$, $m=0,\ldots,M$. The grid points $W_m$, $m=0,1,\ldots,M$, are then given by $W_m=W(0)\exp(X_m)$. The domain $[r_{\min}, r_{\max}]$ is also
 uniformly discretised with step $\delta r=(r_{\max} - r_{\min})/K$ to yield the grid $r_k= r_{\min}+k\delta r$, $k=0,\ldots,K$.

For each grid point $(X_m, r_k)$, the contract value
at time $t_{n-1}^+$ can be expressed as the weighted sum of some contract values at time $t_n^-$. Specifically,  from (\ref{eq_intS}), (\ref{rotation_eq}) and (\ref{eq_qz}) we have
 \begin{equation}\label{eq_qXr}
Q_{t_{n-1}^+}\left(W_m, r_k, A\right) \approx
\frac{P(t_{n-1},t_n)}{\pi}\sum_{i=1,j=1}^{q_1,q_2} \lambda_i^{(q_1)}
\lambda_j^{(q_2)}
Q_{t_n^-}({w_{ijkm}},{r_{ijk}},A),
\end{equation}
\begin{equation}\label{eq_xq}
\begin{split}
{w_{ijkm}}&=\exp\left(\sqrt{2}\tau_x\left(a\xi_i^{(q_1)}+b\xi_j^{(q_2)}\right)+\mu_x(X_m+\ln W(0),r_k)\right),\\
{r_{ijk}}&=\sqrt{2}\tau_r\left(b\xi_i^{(q_1)}+a\xi_j^{(q_2)}\right)+\mu_r(r_k).
\end{split}
\end{equation}
We found that it is more efficient to let $q_1>q_2$. This is because $a>|b|$, i.e. $Z_1$ contributes to
$W$ more than $Z_2$, and thus assigning a larger number of
quadrature points to $Z_1$ is more efficient.

\subsection{Cubic spline interpolations for integration and jump condition}
At time step $(t_{n-1}^+,t_n^-)$, the contract value at
$t=t_n^-$ for any given $A$ is known only at the grid points $(W_m,
r_k)$, $m=0,\ldots,M$, $k=0,\ldots,K$. In order to approximate
the continuous function $Q_{t_n^-}(W,r,\cdot)$ required for the integration, we
propose to  use the bi-cubic spline interpolation over the grid points,
 which is smooth in the first derivative and continuous in the second derivative.
 The error of cubic
spline is $O(h^4)$, where $h$ is the size for the spacing of the
interpolating variable, assuming a uniform spacing. The cubic spline
interpolation involves solving a tri-diagonal system of linear
equations for the second derivatives at all grid points. For a fixed
grid and constant in time model parameters,  the tri-diagonal matrix can be
inverted once and at each time
 step only the back-substitution in the cubic spline procedure is required.
For uniform grids, the bi-cubic spline is about five times as
expensive in terms of computing time as the one-dimensional cubic
spline, as explained below.

 Let $Q^{(X)}_{t}(\cdot)$ denote $Q_{t}(\cdot)$  as a function of $X=\ln(W/W(0))$. Suppose the integration requires the value $Q^{(X)}_{t_n^-}(X,r,\cdot)$ at the point
$(X,r)$ located inside a grid: $X_m\leq X \leq X_{m+1}$ and $r_k\leq
r \leq r_{k+1}$. Because the grid is uniform in both $X$ and $r$,
the second derivatives $\partial^2 Q^{(X)}/\partial X^2$ and $\partial^2
Q^{(X)}/\partial r^2$ can be accurately approximated by the three-point
central difference, and consequently the one-dimensional cubic
spline on a uniform grid involves only four neighboring grid points
for any single interpolation. In our bi-cubic spline case, we can
first obtain $Q^{(X)}$ at four points $(X,r_{k-1})$,
$(X,r_{k})$, $(X,r_{k+1})$, $(X,r_{k+2})$ by applying the
one-dimensional cubic spline on the dimension $X$ for each point and
then we use these four values to obtain $Q^{(X)}(X,r,\cdot)$ through a
one-dimensional cubic spline in $r$. Thus five one-dimensional cubic
spline interpolations are required for a single point, which
involves sixteen grid points neighboring the point of interest
$(X,r)$. Needless to say, this alone will make the pricing of GMWB
under stochastic interest rate much more time consuming than GMWB
under deterministic interest rate.  Not only the evaluation per grid point is more involved but also
evaluation has to be performed for larger number of points.

To apply the jump conditions, only one-dimensional cubic spline
interpolation is involved, as the points before and after each jump
fall on grid points in $r$ and $A$, only interpolation in $X$ is
required.
 Let us introduce an auxiliary finite grid $0 = A_1 < A_2 <
\cdots < A_J = W(0)$ to track the remaining guarantee balance  $A$,
where $J$ is the total number of nodes in the guarantee balance
amount coordinate. The upper limit $W(0)$ is needed because the
remaining guarantee balance cannot exceed the target initial account
value $W(0)$. For each $A_j $, we associate a continuous solution
$Q_t(W,r,A)$ defined by the values at node points $(W_m,r_k)$ and a
two-dimensional cubic spline interpolation over these node points.

 At every jump we let $A$ to be one of
the grid points $A_j ,\;1 \le j \le J$.   Among the infinite number
of possible jumps, a most efficient choice (though not necessary) is
to only allow the guarantee balance to be equal to one of the grid
points $0 = A_1 < A_2 < \cdots < A_J = W(0)$. This implies that, for
a given balance $A_j$ at time $t_n^-$, the possible value after the
withdraw at $t_n^+$ has to be one of the grid points equal to or
less than $A_j$, i.e. $A_j^+=A_i$, $1\leq i \leq j$. In other words,
the withdrawal amount $\gamma$ takes $j$ possible values:
$\gamma=A_j-A_i$, $i=1,\ldots,j$.

Note the above restriction that $\gamma=A_j-A_i$, $i=1,\ldots,j$
is not necessary. The only real restriction is $\gamma \leq A_j$.
However, without the restriction, the value of $A_j^+$ after the jump
falls between the grid points (not exactly on a grid point $A_k$)
and a costly two-dimensional interpolation is required. The error
due to this discretisation restriction can be easily reduced to
acceptable level by increasing $J$.

\begin{figure}[!h]
\begin{center}
\captionsetup{width=0.85\textwidth}
\includegraphics[scale=0.65]{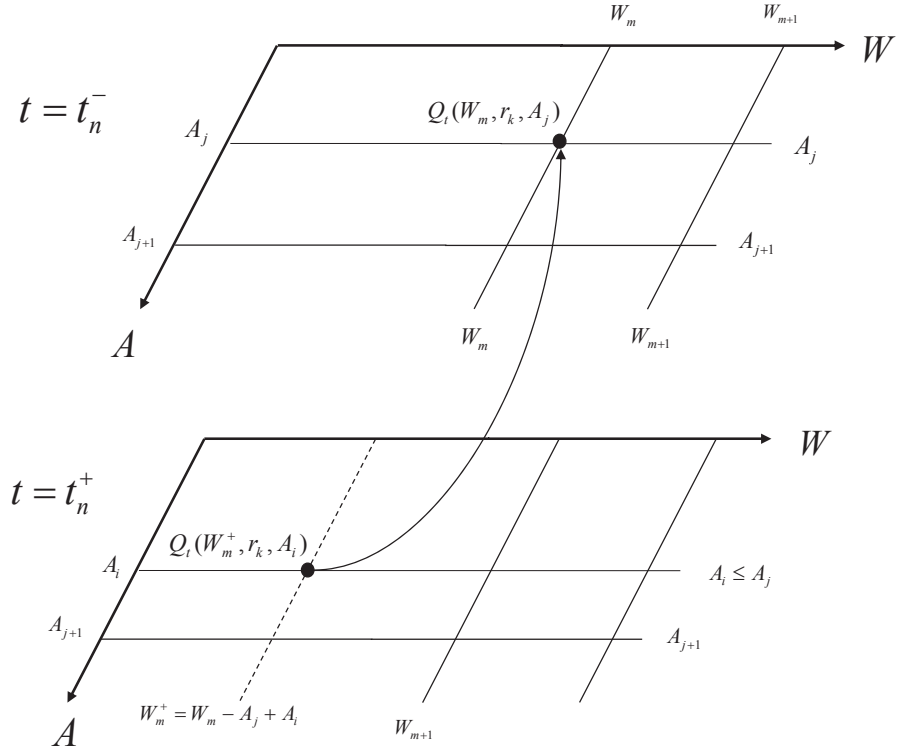}
 \caption{\footnotesize\textbf{Illustration of the jump conditions applied on the finite
difference grids.}} \label{fig1}
\end{center}
\end{figure}

 For any node $(W_m, r_k, A_j)$, $ m=0,1,\ldots, M$, $ k=0,1,\ldots, K$,  $
j=1,\ldots, J$ , given that withdrawal amount can only take the
pre-defined values $\gamma=A_j-A_i$, $i=1,2,\ldots,j$, irrespective
of time $t_n$ and account value $W_m$, the jump condition
(\ref{eqn_jump}) takes the following discrete form
\begin{equation}\label{eqn_jump2}
Q_{t_n^-}(W_m,r_k,A_j)=\max_{1\leq i \leq j}
\left[Q_{t_n^+}(\max(W_m-A_j+A_i,0), r_k,A_i)+C_n(A_j-A_i)\right].
\end{equation}
For optimal strategy, we chose a value for  $1 \leq  i \leq j $
maximizing $Q_{t_n^-}(W_m,r_k,A_j)$. The above jump has to be performed for every node
point $(W_m,r_k,A_j)$, $0\leq m \leq M$,  $ k=0,1,\ldots, K$, $1\leq
j \leq J$ at every withdrawal date. Obviously for every node point
$(W_m,r_k,A_j)$ we have to attempt $j$ jumps to find the maximum
value for $Q_{t_n^-}(W_m,r_k,A_j)$. Figure \ref{fig1} illustrates application of
the  jump condition.

When $W_m-A_j+A_i > 0$, the value $Q_{t_n^+}(W_m-A_j+A_i, r_k,A_i)$
can be obtained by a one-dimensional cubic spline
interpolation from the values at the $M$ discrete
grid points. Interpolation scheme is important, as shown for example in a convergence study by
\citet{Forsyth2002}, it is possible for a PDE based numerical
algorithm for discretely sampled path-dependent option pricing to be
non-convergent (or convergent to an incorrect answer) if the
interpolation scheme is selected inappropriately.

\section{Numerical  Results}\label{NumericalResults_sec}
In this section we first show results for a benchmark test where
closed-form solution exists. We then present numerical results for
pricing GMWB under the static and optimal policyholder strategies using
GHQC algorithm and compare these with the MC and PDE finite
difference results when appropriate.

\subsection{Vanilla European options}
In the case of stochastic dynamics (\ref{referenceportfolio_eq}) for the underlying asset and interest rate, there is a closed-form solution
for European vanilla options, thus providing a valuable benchmark test for numerical algorithms.
In particular, the formulas for prices of vanilla call and vanilla put with strike $K_T$ and maturity $T$, given $S(0)$ and $r(0)$ at time $t_0=0$ are derived in Appendix \ref{bond_vanilla_price_appendix}.

For this test we set the input as follows: asset
volatility $\sigma_S = 20\%$, asset spot value $S(0)=1$, interest
 rate spot value $r(0) = 5\%$, maturity $T=1$, strike $K_T=0.95$, and the
Vasicek interest rate model parameters $\kappa = 0.0349$, $\theta = 5\%$,
$\sigma_r=1\%,\; 3\%$ and $\rho=-0.2, \; 0.0,\; 0.2$.

We calculated the vanilla prices using the finite difference ADI method solving two-dimensional PDE (\ref{eqn_pde}) and
our GHQC method developed in the previous sections and compared with the closed-form solution (\ref{vanilla_closedform_solution}).
Of course in the case of vanilla options, using change of num\'{e}raire to the bond $P(0,T)$, the interest rate dimensionality can
be removed from numerical pricing and the required PDE can be reduced to the one-dimensional PDE similar to deterministic interest rate case.
Here, we implement ADI for the original two-dimensional PDE (\ref{eqn_pde}) for testing and comparison purposes.

For GHQC, we used $q_1=12$ and $q_2=3$ quadrature, i.e. the total number
of quadrature points for each integration is 36. The mesh for GHQC
calculations was fixed at $M=100$ for $X$ dimension and $K=20$ for the interest
rate $r$ dimension, and the total number of time steps $N=5$. Comparing with
typical finite difference calculations, the above mesh and time
steps are quite coarse. Indeed, for ADI calculations, in order to
have a roughly compatible accuracy with GHQC, we had to set $M=200$, $K=40$ and $N=300$.
Table \ref{tab_call} and Table \ref{tab_put} show results for vanilla call
and put prices for different values of $\sigma_r$ and
$\rho$. The percentage numbers in the parentheses in both tables are
the relative numerical errors of the price compared with the
closed-form solution. On average, for $\sigma_r=1\%$ the
relative error
 is about $0.033\%$ for ADI and $0.042\%$ for GHQC,  while  for $\sigma_r=3\%$ the relative error
 is about $0.077\%$ for ADI and $0.040\%$ for GHQC. Significantly, the average relative error for
 ADI is more than doubled when the volatility of interest rate is
 increased from  $\sigma_r=1\%$ to  $\sigma_r=3\%$, while the error for GHQC remains more or less the same.

\begin{table}[!h]\captionsetup{width=0.6\textwidth}
\caption{\footnotesize\textbf{Vanilla call option price for different values
of $\sigma_r$ and $\rho$. The other input parameters are $\sigma =
20\%$, $S(0)=1.0$, $r(0) = 5\%$, $T=1.0$, $K_T=0.95$, $\kappa =
0.0349$ and $\theta = 5\%$.}} \label{tab_call}
\begin{center}
{\footnotesize{\begin{tabular*}{0.6\textwidth}{ccccc} \toprule
 $\sigma_r$ &  $\rho$ &   Closed-form &  ADI  &   GHQC \\
 \midrule
 0.01 & -0.2 & 0.119063  &  0.119096 ($0.027\%$)  &  0.119035 ($0.024\%$) \\
 0.01 &  0.0 & 0.119404  &  0.119461 ($0.027\%$)  &  0.119392 ($0.010\%$) \\
 0.01 &  0.2 & 0.119743  &  0.119775 ($0.026\%$)  &  0.119700 ($0.036\%$) \\
 0.03 & -0.2 & 0.118531  &  0.118604 ($0.062\%$)  &  0.118528 ($0.003\%$) \\
 0.03 &  0.0 & 0.119554  &  0.119626 ($0.060\%$)  &  0.119512 ($0.035\%$) \\
 0.03 &  0.2 & 0.120565  &  0.120634 ($0.057\%$)  &  0.120525 ($0.033\%$) \\
\bottomrule
\end{tabular*}
}}\end{center}
\end{table}
\begin{table}[!h]\captionsetup{width=0.6\textwidth}\caption{\footnotesize\textbf{Vanilla put option price for different values
of $\sigma_r$ and $\rho$. The other input parameters are $\sigma =
20\%$, $S(0)=1.0$, $r(0) = 5\%$, $T=1.0$, $K_T=0.95$, $\kappa =
0.0349$ and $\theta = 5\%$.}} \label{tab_put}
\begin{center}
{\footnotesize{\begin{tabular*}{0.6\textwidth}{ccccc} \toprule
 $\sigma_r$ &  $\rho$ &   Closed-form &  ADI  &   GHQC \\
 \midrule
 0.01 & -0.2 & 0.042547  &  0.042565 ($0.041\%$)  &  0.042522 ($0.059\%$) \\
 0.01 &  0.0 & 0.042888  &  0.042905 ($0.039\%$)  &  0.042878 ($0.024\%$) \\
 0.01 &  0.2 & 0.043227  &  0.043243 ($0.037\%$)  &  0.043187 ($0.093\%$) \\
 0.03 & -0.2 & 0.042132  &  0.042175 ($0.101\%$)   &  0.042133 ($0.002\%$) \\
 0.03 &  0.0 & 0.043156  &  0.043197 ($0.095\%$)  &  0.043117 ($0.091\%$) \\
 0.03 &  0.2 & 0.044167  &  0.044205 ($0.087\%$)  &  0.044133 ($0.077\%$) \\
\bottomrule
\end{tabular*}
}}\end{center}
\end{table}

Both ADI and GHQC took a fraction of a second CPU to calculate one
of the options in Table \ref{tab_call} and Table \ref{tab_put}.
Averaging over 200 calculations for call and put options with the
same inputs as given above, we found the CPU time for each call or
put calculation is 0.055 second for ADI and 0.011 second for GHQC, i.e.
GHQC is about five times faster than ADI in these tests.
All the calculations shown in this study were performed on a desktop
 with an Intel Core i5-4590 CPU@3.30GHz with a 4.00GB RAM.

It is worth commenting that for the vanilla call and put options the
final payoff function is only piecewise linear, i.e. it is not  a
polynomial function and it is not smooth at the strike $K_T$ (first
derivative discontinuous at $W=K_T$). If we apply the
Gauss-Hermite quadrature
 only to the half domain ($W\geq K_T$ for call and $W\leq K_T $
 for put), then the payoff function at the maturity is a simple linear function needing no interpolation anywhere and
  we found the GHQC calculations can be made as
 accurate  as desired, virtually limited only by the machine accuracy. In the
 above tests we did not take advantage of this specific feature
 because it is not generally applicable.

\subsection{GMWB pricing results}
In the case of the static policyholder behavior, the withdrawal amounts are predetermined
at the beginning of the contract. In this case the paths of the wealth account $W(t)$ can be simulated
 and a standard MC simulation method can be used to calculate the price of VA contract with GMWB.
 Below we show results for both static and
  optimal withdrawal cases. The static case allows a comparison between MC and GHQC, further validating
  the new algorithm. We  have also implemented an efficient finite difference
 ADI algorithm for pricing of VA with GMWB under the static
policyholder behavior solving corresponding two-dimensional PDE.  In what follows, results from the GHQC method
 will be compared with the MC and ADI methods when applicable.

\subsubsection{GMWB with static policyholder behavior}
In a static case the withdrawal amount is pre-determined for each
withdrawal date. In this case  at each payment date  the jump
condition applies to the single solution (therefore no need for a
grid in the guarantee account $A$ dimension). Since the withdrawal amount is known
at every payment date, the stochastic paths of the underlying $W$
can be simulated by MC method. Here we compare GHQC results
with those of MC and ADI methods.

Table \ref{tab_static_Rho} shows the prices of the VA with GMWB under the static withdrawal strategy as a function
of the correlation $\rho$, comparing results between MC,
ADI and GHQC.
 The model input parameters are $\sigma_S = 20\%$, $\sigma_r =
 2\%$, $S(0)=W(0)=1.0$, $r(0) = 5\%$, $g=10\%$, $T=1/g=10.0$, $N_w=4$
(quarterly withdraw frequency), $\alpha=0.006$, $\kappa = 0.0349$
and $\theta = 5\%$. In \cite{Kwok2012StochInterestGMWB}, results for static GMWB pricing
 using similar parameters  were presented  for the lower and upper bound
estimates of the price comparing with the MC simulation results. However, those results
 are for a continuous in time withdrawal case, while our results are for the discrete in time withdrawals at a quarterly frequency.

For the MC method, we have used one million sample paths simulated from the
closed-form transition density under the new measure
$\widetilde{\mathbb{Q}}$, so that there is no time discretization error. For both ADI and GHQC we have used two
meshes, one coarser and one finer, with the finer mesh doubles the
number of node points in both $r$ and $X$ dimensions. Let $M_A$ denote the
coarser mesh for ADI with $M=200$ for $X$ dimension, $K=50$ for $r$ dimension, and
$M_A^\star$ denote finer mesh for ADI with $M=400$ and $K=100$. In
addition, the number of time steps for each period (between
consecutive withdrawal dates) was set $N_{\Delta t}=100$ for ADI
when using the coarser mesh, and it was doubled to $N_{\Delta
t}=200$ when using the finer mesh $M_A^\star$. For GHQC, the coarser
mesh, denoted as $M_G$, has $M=50$ and $K=30$, and the finer mesh
for GHQC, denoted as $M_G^\star$, has $M=100$ and $K=60$. For both
$M_G$ and $M_G^\star$ meshes, we have used a single time step
between withdrawal dates, i.e. $N_{\Delta t}=1$. For the quadrature
points, we used $q_1=5$ and $q_2=3$ for the coarser mesh, and
$q_1=9$ and $q_2=5$  for the finer mesh. Note the finer mesh
$M_G^\star$ for GHQC is overall much coarser than the coarser mesh
$M_A$ for ADI. Comparing with typical finite difference PDE calculations required for pricing financial derivatives, the finer mesh $M_G^\star$ is actually very coarse.

In Table \ref{tab_static_Rho}, the numbers in the parentheses next to
the MC results are the standard errors due to the finite number of simulations, while
for ADI and GHQC the numbers in the parentheses are the relative
difference from MC results. On
average, the relative standard error for MC (standard error divided
by the estimated mean) is 4.6E-4, sufficiently small for the MC
results to serve as a basis to compare among different results\footnote{Hereafter, $a$E-$b$ denotes $a\times 10^{-b}$.}.

 If a set of numerical results have the same or better accuracy than the
MC results, then the relative difference between this set of results
and MC results should be in the same order of magnitude as the
relative standard error of the MC. Results in Table
\ref{tab_static_Rho} show that, the average relative difference
between ADI and MC is about 9.2E-4 for the coarser mesh $M_A$, and
it is about 6.9E-4 for the finer mesh $M_A^\star$. In comparison,
the average relative difference between GHQC and MC is about 5.6E-4
for the coarser mesh $M_G$, and it is about 3.7E-4 for the finer
mesh $M_G^\star$. These relative differences are indicative that the
GHQC calculations are perhaps more accurate than the ADI, even
comparing GHQC results of coarser mesh $M_G$ with ADI results of the
finer mesh $M_A^\star$.

\begin{table}[!h]\captionsetup{width=0.9\textwidth}\caption{\footnotesize\textbf{Price of VA with GMWB under static withdrawal strategy for different values
 $\rho$. Other parameters: $\sigma = 20\%$,  $\sigma_r =
 2\%$,
$S(0)=1.0$, $r(0) = 5\%$, $g=10\%$ ($T=1/g$), $N_w=4$, $\alpha=0.006$, $\kappa = 0.0349$, $\theta = 5\%$.}} \label{tab_static_Rho}
\begin{center}
{\footnotesize{\begin{tabular*}{0.9\textwidth}{cccccc} \toprule
 $\rho$ &   MC &  ADI ($M_A$) &   ADI ($M_A^\star$) &   GHQC ($M_G$)  &   GHQC ($M_G^\star$) \\
 \midrule
-0.6 & 1.004826 (3.1E-4) & 1.00616 (1.3E-3) & 1.00532 (4.9E-4) &
1.00557 (7.4E-4) & 1.00484 (1.4E-5)\\
-0.4 & 1.011952 (4.5E-4) & 1.01338 (1.4E-3) & 1.01255 (5.9E-4) &
1.01295 (9.8E-4) & 1.01236 (4.0E-4) \\
-0.2 &  1.019002 (4.8E-4) & 1.02024 (1.2E-3) & 1.01942 (4.1E-4) &
1.01982 (8.0E-4) & 1.01945 (4.4E-4) \\
0.0 &  1.026177 (4.8E-4) & 1.02675 (5.6E-4) & 1.02592 (2.5E-4) & 1.02625 (7.9E-5) &  1.02613 (4.6E-5) \\
0.2 & 1.032256 (4.8E-4) & 1.03289 (6.1E-4) & 1.03206 (1.9E-4) & 1.03279 (5.2E-4) & 1.03249 (2.3E-4)  \\
0.4 & 1.038966 (5.3E-4) & 1.03867 (2.8E-4) & 1.03784 (1.1E-3) &
1.03886 (1.0E-4) & 1.03849 (4.6E-4) \\
0.6 & 1.045171 (5.8E-4) & 1.04407 (1.1E-3) & 1.04325 (1.8E-3) &
1.04445 (6.9E-4) & 1.04413 (9.9E-4) \\
\bottomrule
\end{tabular*}
}}\end{center}
\end{table}

Not only the GHQC may have better accuracy than the ADI, the CPU
time comparison is even more impressive: for ADI calculations, the
CPU time per price is 1.0 second and 7.6 second for the coarser and
finer mesh, respectively; while for GHQC calculations, the CPU time
per price is 0.02 second and 0.24 second for the coarser and finer
mesh, respectively. In other words, comparing the coarser mesh
calculations, the GHQC is about 50 times as fast as ADI, and
comparing the finer mesh calculations the GHQC is about 30 times as
fast as ADI, while achieving better accuracy. Only the GHQC results
 with the finer mesh has an average  relative difference (relative to MC
 results) smaller than the average relative standard error of MC with one million simulations.

\begin{table}[!h]\captionsetup{width=0.8\textwidth}\caption{\footnotesize\textbf{Prices of VA with GMWB under static withdrawal strategy for different fees
 $\alpha$ and positive correlation   $\rho=0.3$. Other parameters: $\sigma = 20\%$,  $\sigma_r =
 2\%$,
$S(0)=1.0$, $r(0) = 5\%$, $g=10\%$ (i.e. $T=1/g$), $N_w=4$, $\kappa = 0.0349$ and $\theta =
5\%$.}} \label{tab_static_a1}
\begin{center}
{\footnotesize{\begin{tabular*}{0.6\textwidth}{cccc} \toprule
 $\alpha$ &   MC &    ADI ($M_A^\star$) &   GHQC ($M_G^\star$) \\
 \midrule
0  & 1.064589 (5.2E-4) &1.06389 (6.6E-4) & 1.06434 (2.3E-4) \\
25 & 1.052354 (5.0E-4) &1.05154 (7.7E-4)  &1.05202 (3.2E-4) \\
50 & 1.040172 (4.9E-4) &1.03963 (5.2E-4)  &1.04015 (2.2E-5) \\
75 & 1.029112 (4.9E-4) &1.02817 (9.2E-4)  &1.02873 (3.7E-4) \\
100 &1.018198 (4.7E-4) & 1.01714 (1.0E-3) & 1.01773 (4.6E-4) \\
125 &1.007269 (4.7E-4) &1.00653 (7.3E-4) & 1.00716 (1.1E-4) \\
150 &0.997382 (4.5E-4) & 0.996316 (1.1E-3) & 0.996993  (3.9E-4) \\
175 &0.987463 (4.4E-4) &0.986501 (9.7E-4) &0.987222    (2.4E-4)\\
200 &0.977950 (4.3E-4) &0.977069 (9.0E-4) &0.977835    (1.2E-4) \\
\bottomrule
\end{tabular*}
}}\end{center}
\end{table}
\begin{table}[!h]\captionsetup{width=0.8\textwidth}\caption{\footnotesize\textbf{Prices of VA with GMWB under static withdrawal strategy for different fees
 $\alpha$ and negative correlation   $\rho=-0.3$. Other parameters: $\sigma = 20\%$,  $\sigma_r =
 2\%$,
$S(0)=1.0$, $r(0) = 5\%$, $g=10\%$ (i.e. $T=1/g$), $N_w=4$, $\kappa = 0.0349$ and $\theta =
5\%$.}} \label{tab_static_a2}
\begin{center}
{\footnotesize{\begin{tabular*}{0.6\textwidth}{cccc} \toprule
 $\alpha$ &   MC & ADI ($M_A^\star$) &   GHQC ($M_G^\star$) \\
 \midrule
0  & 1.044794 (5.3E-4) &1.04526 (4.5E-4)   &1.04495   (1.5E-4) \\
25 & 1.032550 (5.1E-4) &1.03274 (1.8E-4)   &1.03253   (2.0E-5) \\
50 & 1.020225 (5.0E-4) &1.02071 (4.7E-4)   &1.02059   (3.6E-4) \\
75 & 1.009109 (5.0E-4) &1.00915 (4.1E-5)   &1.00912   (1.1E-5)\\
100& 0.9978363 (4.8E-4)& 0.998039 (2.0E-4) &0.998104   (2.7E-4) \\
125& 0.9871583 (4.8E-4)& 0.987377 (2.2E-4) &0.987531   (3.8E-4) \\
150& 0.9770732 (4.6E-4)& 0.977148 (7.7E-5) &0.977387   (3.2E-4) \\
175& 0.9673112 (4.4E-4)& 0.967338 (2.8E-5) &0.967662   (3.6E-4) \\
200& 0.9581683 (4.4E-4)& 0.957938 (2.4E-4) &0.958343   (1.8E-4) \\
\bottomrule
\end{tabular*}
}}\end{center}
\end{table}

Table \ref{tab_static_a1} shows the price of the VA with GMWB under the static withdrawal strategy as a function
of the fee $\alpha$ in the unit of basis point (a basis point is 0.01\%
). The correlation $\rho$ is fixed at $\rho=0.3$ and
all the other inputs are the same as for the calculations for Table
\ref{tab_static_Rho}. For ADI and GHQC, only results of the finer
meshes $M_A^\star$ and $M_G^\star$ are shown. The number of
simulations for MC and the number of time steps for ADI and GHQC
were unchanged. In this case the average relative standard error for
MC is 4.7E-4. The average relative difference between ADI and MC is
8.4E-4, somewhat larger than the MC standard error. The average
relative difference between GHQC and MC is 2.5E-4, which is about
half of the MC standard error.

Table \ref{tab_static_a2} shows the same results as Table
\ref{tab_static_a1} except the correlation is negative at
$\rho=-0.3$. In this case the average relative standard error for MC
is 4.8E-4. The average relative difference between ADI and MC is
2.1E-4, and the average relative difference between GHQC and MC is
2.3E-4, both are about half of the MC standard error.

It is interesting to compare the results of Vasicek model
with the case of deterministic interest rate set to be $r(0)$ during the contract life. We found that, for the test cases shown in Table
\ref{tab_static_a1} where the correlation is positive at $\rho=0.3$,
the static withdrawal strategy GMWB prices under the stochastic interest rate are about
$2\%$ larger than the deterministic counterpart, i.e. the ratio in
prices is about $1.02$. However, a small difference in the
price for a given fee does not not mean a small difference
 in the fair fees given the premium. A fair fee $\alpha$ is the fee making the initial premium
  equal to the contract price, i.e. $Q(W(0),r(0),W(0)) = W(0) $.
   For the inputs given for Table \ref{tab_static_a1}, we found the
   fair fee for GMWB under the stochastic interest rate
 is 143 basis points, which is $49\%$ higher than the deterministic
 case of 95.8 basis points.

 On the other hand, for the cases in Table
\ref{tab_static_a2} where the correlation is negative at
$\rho=-0.3$, the prices of VA with the static withdrawal GMWB under the stochastic interest
rate are virtually the same as the deterministic interest rate counterpart -- on
average the relative difference in prices is about 3.5E-4, which has
the same magnitude as the average relative errors. The corresponding
relative difference
 in the fair fees of GMWB between the stochastic interest rate and deterministic rate cases is
only about $0.2\%$, which could be in the same order of magnitude as
 numerical errors in the fees.

\subsubsection{GMWB under optimal withdrawals}
Having numerically validated the implementation of GHQC algorithm,
we then proceed to perform calculations for pricing of GMWB under the dynamic withdrawal strategy. Note that
for the dynamic strategy GMWB pricing, exactly the same numerical functions are used as
for the static strategy GMWB case. The only extra step required for the
dynamic case is simply finding the optimal amount among possible withdrawal values, while in the static case only the
fixed withdrawal amount is considered. In particular, the integration
and jump condition application all use identical functions in
the dynamic and static cases.

Table \ref{tab_dynamic1} is the dynamic strategy counterpart of Table
\ref{tab_static_a1}, i.e. all the inputs (model parameters, contract
details, mesh, quadrature points and time step settings) are
the same, but the calculations are for the dynamic withdrawals. In
this example the number of grid points in the guarantee account $A$
is $J=100$. An extra input needed for the dynamic case is the
penalty coefficient $\beta$, which is fixed at $\beta=10\%$ in all
the following calculations. A simple extra validation for the
dynamic calculations is to set the penalty coefficient $\beta$ very
high, say  $\beta=50\%$, then the price under the optimal
withdrawals should be the same as under the static withdrawals, which was indeed
confirmed by our numerical tests.

\begin{table}[!h]\captionsetup{width=0.82\textwidth}\caption{\footnotesize\textbf{Prices of VA with GMWB under dynamic withdrawal strategy for different fees
 $\alpha$ in the case of negative and positive correlation $\rho$. Other parameters: $\sigma_S = 20\%$,  $\sigma_r =
 2\%$,
$S(0)=1.0$, $r(0) = 5\%$, $g=10\%$ ($T=1/g$), $\beta=10\%$,
$N_w=4$, $\kappa = 0.0349$,
$\theta = 5\%$.}} \label{tab_dynamic1}
\begin{center}
{\footnotesize{\begin{tabular*}{0.4\textwidth}{ccc} \toprule
 $\alpha$ &     GHQC ($\rho=-0.3$) &   GHQC ($\rho=0.3$) \\
 \midrule
0   & 1.08348 & 1.10173 \\
25  & 1.06651 & 1.08367 \\
50  & 1.05107 & 1.06707 \\
75  & 1.03719 & 1.05184 \\
100 & 1.02484 & 1.03804 \\
125 & 1.01389 & 1.02558 \\
150 & 1.00408 & 1.01446 \\
175 & 0.995356 & 1.00463 \\
200 & 0.987673& 0.996057 \\
\bottomrule
\end{tabular*}
}}\end{center}
\end{table}

\begin{figure}[!h]
\captionsetup{width=0.8\textwidth}
\begin{center}
\includegraphics[scale=0.75]{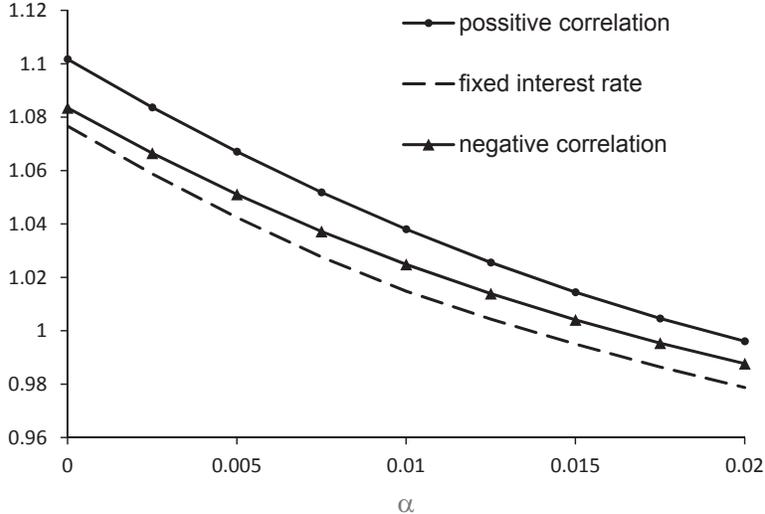}
 \caption{\footnotesize\textbf{Prices of VA with GMWB under optimal withdrawals and  stochastic interest rate with $\rho=0.3$ and $\rho=-0.3$. The other
  parameters are the same as for Table \ref{tab_dynamic1}. }} \label{fig2}
\end{center}
\end{figure}

  We also show comparison of the
prices for the cases of deterministic and stochastic interest rates
in Figure \ref{fig2}. Similar to the static withdrawal case, with a positive
correlation between asset and interest rate $\rho=0.3$, prices of the VA with dynamic withdrawal GMWB
 under the stochastic interest rate are about $2\%$ larger
than in the case of deterministic constant interest rate
set to $r(0)$, i.e. the ratio in prices is about $1.02$. Again, a small
difference in the price for a given fee does not not mean a
small difference in the fair fees given the premium.
   For the inputs given for Table \ref{tab_dynamic1}, we found the
   fair fee for GMWB under the stochastic interest rate
 is 188 basis points, which is $38\%$ higher than the fair fee 136 basis points in the case of deterministic
 interest rate.

Similar to the static withdrawal case, in the case of negative
correlation between asset and interest rate $\rho=-0.3$, the differences of
prices between cases under the stochastic interest rate
and cases under a deterministic interest rate $r(0)$ are
relatively small, compared with the case of positive correlation.
Now, at $\rho=-0.3$,  these differences are only about $0.9\%$ on
average, much smaller than $2\%$ when the correlation is at
$\rho=0.3$. However, unlike the static withdrawal case where for $\rho=-0.3$
the difference of prices between stochastic rate and deterministic
rate is negligible (as small as the numerical errors), the
difference of $0.9\%$ is still
 significant (the estimated relative numerical error is in the order of $0.01\%$), leading to a  significant difference in the fair
 fees. We found for this case at $\rho=-0.3$, the fair fee under the
 stochastic rate is 161 basis points, which is about $19\%$ higher
 than the deterministic interest rate counterpart (136 basis points).

%
%

As expected, the CPU time for these calculations of dynamic withdrawal cases is
much longer than the static withdrawal counterpart with the same mesh and time
steps. The average CPU time for  calculating the prices in Table
\ref{tab_dynamic1} is about 39 seconds per price, comparing with
0.24 second per price for the static withdrawal case with the same mesh and
time steps. The ratio of these CPUs is about 160, which is
reasonable because instead of tracking a single
two dimensional solution in $(X,r)$  for the static withdrawal case, for the
dynamic withdrawal case we have to track $J=100$ such solutions for all the
nodes in $A$. In addition, on each withdrawal date all the
possible jumps have to be performed in order to find the optimal
withdrawal amount for each grid point in $(X,r,A)$ space, as compared to the
static withdrawal case where only one jump is needed for each single grid
 point in $(X,r)$ space.

  \section{Conclusion}\label{conclusion_sec}
   In this
paper we developed a new  direct integration method for pricing of the VA with guarantees under the both static and dynamic (optimal)
policyholder behaviors in the case of stochastic interest rate. Using bond
price as a num\'{e}raire,  we have derived the closed-form bivariate
transition density for the correlated state variables $\ln S(t)$ and
$r(t)$. Under the new measure the required expectations are reduced
to the two-dimensional integrals which can be readily calculated through
the two-dimensional numerical integration using Gauss-Hermite
quadrature, allowing an efficient backward time stepping for solving
recursive Bellman equation. A spectral rotation scheme
preserves the symmetry with respect to the principal axes of the
bivariate random field, yielding a robust and accurate quadrature
application. A two-dimensional cubic spline interpolation on the
finite grids for  $(X(t), r(t))$ is utilized to provide the
continuous function required by the quadratures.
  The proper jump conditions are applied at each withdrawal date that allows the optimal withdrawal
 decision to be made.

The algorithm is convincingly validated by comparing to the European option
pricing where closed-form analytical solution exists, and by static withdrawal
GMWB pricing where MC and PDE ADI methods can provide good benchmark
solutions. Numerical tests show the accuracy of the presented GHQC
method is at least compatible to a typical ADI finite difference
scheme, but it is more robust and significantly faster.

For dynamic withdrawal GMWB pricing, using the new algorithm we found some
interesting results which we believe are new to the literature. When the correlation between the underlying risky asset and
interest rate is positive, the GMWB price or equivalently the fee
under the stochastic interest rate is significantly higher than in the case of
deterministic interest rate. In the particular test
problem,  the fee in the case of
stochastic interest rate  is about $40\%$ higher than the deterministic case
when $\rho=0.3$. When the correlation is negative, the differences
are still significant but much less than  in the case of positive correlation. The fee in the case of stochastic interest rate  is about $20\%$ higher
than the deterministic case when correlation is negative at
$\rho=-0.3$. On the other hand, the situation in the static withdrawal pricing
is remarkably different: at negative correlation, the differences in prices and
fees of GMWB between stochastic and deterministic settings are
virtually negligible.

In this paper we focused on pricing of a very basic GMWB structure. However, presented algorithm can be easily adapted to pricing other VA guarantees and solving similar stochastic control problems with two state variables possibly affected by control. Applications to pricing Asian, barrier and other financial derivatives
with a single underlying risky and stochastic interest rate are straightforward. Also, it should be possible to extend the algorithm to situations when the underlying bivariate transition density is not known in closed-form but its moments are known, similarly as developed in \cite{LuoShevchenkoGHQC2014} for one-dimensional problems; this is a subject of future research.

\section{Acknowledgement} This
research was supported by the CSIRO-Monash Superannuation Research
Cluster, a collaboration among CSIRO, Monash University, Griffith
University, the University of Western Australia, the University of
Warwick, and stakeholders of the retirement system in the interest
of better outcomes for all. This research was also partially supported under the
Australian Research Council's Discovery Projects funding scheme (project number:
DP160103489).


\appendix

\section{Joint Distribution of $S(t)$, $r(t)$ and $Y(t)=\int_{0}^{t}r(u)du$}\label{appendixA}
Consider the probability measure $\mathbb{Q}$ with corresponding stochastic processes for $S(t)$ and $r(t)$ given by (\ref{referenceportfolio_eq}), and the new probability measure $\widetilde{\mathbb{Q}}$ obtained via the Radon-Nikodym derivative
 \begin{equation}
 \mathbb{Z}_t=\left.\frac{d\widetilde{\mathbb{Q}}}{d\mathbb{Q}}\right|_{\mathcal{F}_t}=\frac{M(0)}{M(t)}\frac{P(t,T)}{P(0,T)},\quad t\in [0,T].
 \end{equation}
Here, $M(t)=e^{\int_0^t r(\tau)d\tau}$ is the money market account and $P(t,T)=\mathrm{E}^{\mathbb{Q}}_0[e^{-\int_t^T r(\tau)d\tau}]$ is the $T$-maturity bond. In particular, it is easy to see that $\mathbb{Z}_T>0$ and $\mathrm{E}_0^\mathbb{Q}[\mathbb{Z}_T]=1$ and this change of measure leads to the following formula for an arbitrary function $f(S(T),r(T))$
$$
\mathrm{E}^{\mathbb{Q}}_{0}\left[e^{-\int_0^T r(u)du}f(S(T),r(T))\right] =P(0,T){\mathrm{E}}^{\widetilde{\mathbb{Q}}}_{0}\left[f(S(T),r(T)\right],
$$
assuming that these expectations exist. One can say that we changed num\'{e}raire from the money market account $M(t)$ to the $T$-maturity bond $P(t,T)$.
 Using It\^{o}'s formula, the process for $\mathbb{Z}_t$ is easily obtained from the process (\ref{bond_sde}) for the bond price $P(t,T)$ to be
 $$
 d\mathbb{Z}_t=\phi(t)\mathbb{Z}_t d\mathcal{B}_1,\quad \phi(t)=-\sigma_r B_{t,T},
 $$
 where $B_{t,T}=(1-e^{-\kappa (T-t)})/\kappa$, and Girsanov theorem gives the corresponding transformation to the Wiener process $\mathcal{B}_1(t)=\phi(t)dt+d\widetilde{\mathcal{B}}_1(t)$. Thus the processes for $S(t)$ and $r(t)$ under the new measure $\widetilde{\mathbb{Q}}$ for $t\in [0,T]$ are
 \begin{equation}\label{new_riskneutralprocess2}
 \begin{split}
dS(t)/S(t)&=(r(t)+\sigma_S \rho \phi(t)) dt +\sigma_S \left(\rho d\widetilde{\mathcal{B}}_1(t) +\sqrt{1-\rho^2} d\widetilde{\mathcal{B}}_2(t)\right),\\
dr(t)&=\kappa\left(\widetilde\theta(t)-r(t)\right)dt+\sigma_r d\widetilde{\mathcal{B}}_1(t);\;\widetilde\theta(t)=\theta+\frac{\sigma_r}{\kappa}\phi(t)
\end{split}
\end{equation}
with $\widetilde{\mathcal{B}}_1(t)$ and $\widetilde{\mathcal{B}}_2(t)$ independent Wiener processes.

In this section we derive the joint Normal distribution of $\ln S(t)$ and $r(t)$ for given $S(0)$ and $r(0)$ under the new probability measure $\widetilde{\mathbb{Q}}$. We also derive the bond price, European vanilla price formulas, and 3d joint Normal distribution of $\ln  S(t)$, $r(t)$ and $Y(t)=\int_{0}^{t}r(u)du$ conditional on $S(0)$ and $r(0)$ under the measure $\mathbb{Q}$. The last is useful for validation tests to simulate and calculate the contract payoff without time discretization error. Some of these formulas can be found in the literature, e.g. see \cite[appendix B1 and section 4.5]{Cairns2004} for bond price and distribution of $(r(t),Y(t))$ under the Vasicek model, but are presented here for completeness and notational consistency.

 The formulas derived below for the mean and covariances can be used for simulation of $(S({t_n}), r({t_n}))$ given $(S({t_{n-1}}), r({t_{n-1}}))$. One has to just set $t\rightarrow t_n-t_{n-1}$, $T\rightarrow T-t_{n-1}$, and $r(t_0)\rightarrow r({t_{n-1}})$, $S(0)\rightarrow S({t_{n-1}})$ in these formulas. To obtain the mean and covariance formulas (\ref{mean_cov_eq}) required to calculate expectation (\ref{optionprice_new}) over $(t_{n-1},t_n)$ one has to set $t=T\rightarrow t_n-t_{n-1}$ and subtract  $\alpha \times (t_n-t_{n-1})$ from the mean of $\ln S(t_n)$ to get the mean of $\ln W(t_n)$.

\subsection{Distribution for $r(t)$}
The solution for the interest rate $r(t)$ given $r(0)$, with the process in (\ref{new_riskneutralprocess}) under $\widetilde{\mathbb{Q}}$, is
\begin{equation}\label{r_new_solution}
r(t)=r(0) e^{-\kappa t}+e^{-\kappa t}\kappa\int_0^t\widetilde\theta(\tau)e^{\kappa\tau}d\tau +\sigma_r e^{-\kappa t} \int_0^t e^{\kappa\tau} d\widetilde{\mathcal{B}}_1(\tau)
\end{equation}
 that can be checked directly by denoting $H_t=\int_0^t e^{\kappa\tau} d\widetilde{\mathcal{B}}_1(\tau)$, $r(t):=g(t,H_t)$ and then calculating $dr(t)=dg(t,H_t)$ using It\^{o} formula to obtain the process $dr(t)$ in (\ref{new_riskneutralprocess}). Thus, $r(t)$ conditional on $r(0)$ is from Normal distribution with
\begin{eqnarray}\label{mean_var_r_eq}
\begin{split}
\mu_r(t):=\mathrm{mean}(r(t))&=r(0) e^{-\kappa t}+e^{-\kappa t}\kappa\int_0^t \widetilde\theta(\tau)e^{\kappa\tau}d\tau,\\
\mathrm{var}(r(t))&=\frac{\sigma_r^2}{2\kappa}\left(1-e^{-2\kappa t}\right).
\end{split}
\end{eqnarray}
In the case of constant in time parameter $\theta$, simple integration yields
$$
\mu_r(t)=r(0) e^{-\kappa t}+\theta\left(1-e^{-\kappa t}\right)+\frac{\sigma_r^2}{2\kappa^2}\left((1-e^{-2\kappa t}) e^{-\kappa (T-t)}-2(1-e^{-\kappa t}) \right).
$$
In the case of risk-neutral process (\ref{referenceportfolio_eq}) under the measure $\mathbb{Q}$, the last term with factor ${\sigma_r^2}$  in the above formula for $\mu_r(t)$ should be set to zero and no change to the variance is required.

\subsection{Distribution for $Y(t)=\int_{0}^{t}r(u)du$}\label{distribution_Yt_sec}
Using solution (\ref{r_new_solution}) for $r(t)$, direct calculation of $Y(t)$ under the probability measure $\widetilde{\mathbb{Q}}$ gives
\begin{eqnarray}\label{stochinterest_integral_solution}
Y(t)=\int_0^t r(\tau) d\tau&=&\int_0^t \mu_r(\tau)d\tau+\sigma_r\int_0^t e^{-\kappa\tau} d\tau \int_0^\tau e^{\kappa s} d\widetilde{\mathcal{B}}_1(s) ds\nonumber \\
&=&\int_0^t \mu_r(\tau)d\tau+\sigma_r\int_0^t d\widetilde{\mathcal{B}}_1(s)\int_s^t e^{-\kappa(\tau-s)}d\tau\nonumber \\
&=&\int_0^t \mu_r(\tau)d\tau+\frac{\sigma_r}{\kappa}\int_0^t (1-e^{-\kappa(t-s)})d\widetilde{\mathcal{B}}_1(s),
\end{eqnarray}
where the $2d$ integral involving $\widetilde{\mathcal{B}}_1(t)$ was simplified by changing order of the integrations. Thus the distribution of $Y(t)$ is Normal with the mean and variance calculated via the standard integrations
\begin{eqnarray}\label{mean_var_Yt_eq}
\begin{split}
I_1(t):=\mathrm{mean}(Y(t))&=\int_0^t \mu_r(\tau)d\tau,\\
\mathrm{var}(Y(t))&=\frac{\sigma_r^2}{\kappa^2}\int_0^t \left(1-e^{-\kappa(t-s)}\right)^2 ds\nonumber \\
&=\frac{\sigma_r^2}{\kappa^2}\left(t-\frac{2}{\kappa}(1-e^{-\kappa t})+\frac{1}{2\kappa}(1-e^{-2\kappa t})\right),
\end{split}
\end{eqnarray}
where $I_1(t)$ in the case of constant in time parameter $\theta$ can be found in closed-form
$$I_1(t)=\frac{1}{\kappa}(1-e^{-\kappa t})\left(r_0-\theta+\frac{\sigma_r^2}{2\kappa^2}\left(2-e^{-\kappa T}+e^{-\kappa(T-t)}\right)\right)+
\left(\theta-\frac{\sigma_r^2}{\kappa^2}\right)t.
$$
In the case of the risk-neutral process (\ref{referenceportfolio_eq}) under the measure $\mathbb{Q}$, $\sigma_r$ in the above formula for $I_1(t)$ should be set to zero and no change to the variance is required.

\subsection{Distribution for $\ln S(t)$}\label{appendix_Spdf}
The solution for $\ln S(t)$ given $\ln S(0)$, with the process in (\ref{new_riskneutralprocess}) under $\widetilde{\mathbb{Q}}$, is given by
\begin{equation}\label{xsolution_eq}
\begin{split}
\ln S(t)=&\ln S(0)+\int_0^t r(\tau) d\tau +\rho\sigma_S\int_0^t \phi(\tau)d\tau-\frac{1}{2}\sigma_S^2 t\\
         &+\rho\sigma_S\int_0^t d\widetilde{\mathcal{B}}_1(\tau)+\sigma_S\sqrt{1-\rho^2}\int_0^t d\widetilde{\mathcal{B}}_2(\tau).
\end{split}
\end{equation}
Substituting (\ref{stochinterest_integral_solution}) it is easy to see that $\ln S(t)$ is from Normal distribution and performing simple integrations obtain
\begin{eqnarray}\label{mean_var_X_eq}
\begin{split}
\mathrm{mean}(\ln S(t))&=\ln S(0)+I_1(t)-\frac{1}{2}\sigma_S^2 t- \frac{\rho \sigma_S \sigma_r}{\kappa^2}\left(\kappa t- e^{-\kappa T}(e^{\kappa t}-1)\right),\\
\mathrm{var}(\ln S(t))&=\sigma_S^2 t+\frac{\sigma_r^2}{2\kappa^3}\left(2\kappa t-3+4e^{-\kappa t}-e^{-2\kappa t}   \right)+\frac{2\rho\sigma_S\sigma_r}{\kappa^2}\left(\kappa t-1+e^{-\kappa t}\right).\nonumber
\end{split}
\end{eqnarray}
In the case of risk-neutral process (\ref{referenceportfolio_eq}) under the measure $\mathbb{Q}$, the last term proprotional to $\rho$ in the above formula for $\mathrm{mean}(\ln S(t))$ should be set to  zero and no change to the variance is required.

\subsection{Covariances between $\ln S(t)$, $r(t)$ and $Y(t)=\int_{0}^{t}r(u)du$}
Direct calculations using solutions (\ref{xsolution_eq}), (\ref{r_new_solution}) and (\ref{stochinterest_integral_solution}) for $\ln S(t)$, $r(t)$ and $Y(t)$ conditional on $\ln S(0)$ and $r(0)$, under the probability measure $\widetilde{\mathbb{Q}}$, yield
\begin{eqnarray}\label{cov_xr_eq}
\begin{split}
\mathrm{cov}(\ln S(t),r(t))&=\sigma_r\int_0^t \left(\sigma_S\rho+\frac{\sigma_r}{\kappa} \left(1-e^{-\kappa(t-s)}\right)\right) e^{-\kappa (t-s)}ds\nonumber \\
&= \frac{\rho\sigma_S\sigma_r}{\kappa}\left( 1-e^{-\kappa t}\right)+\frac{\sigma_r^2}{2\kappa^2}\left(
1-2e^{-\kappa t}+e^{-2\kappa t}\right),\\
\mathrm{cov}(Y(t),r(t))&=\frac{\sigma_r^2}{\kappa} e^{-\kappa t} \int_0^t e^{\kappa s} \left(1-e^{-\kappa(t-s)}\right) ds\nonumber\\
&=\frac{\sigma_r^2}{2\kappa^2} \left(1-2e^{-\kappa t}+e^{-2\kappa t}\right),\\
\mathrm{cov}(\ln S(t),Y(t))&=\frac{\sigma_r}{\kappa}\int_0^t (1-e^{-\kappa(t-s)}) \left(\sigma_S\rho+\frac{\sigma_r}{\kappa} \left(1-e^{-\kappa(t-s)}\right)\right)ds\nonumber\\
&=\frac{\rho\sigma_S\sigma_r}{\kappa^2}\left(\kappa t-1+e^{-\kappa t}\right)+\frac{\sigma_r^2}{2\kappa^3}\left(2\kappa t-3+4e^{-\kappa t}-e^{-2\kappa t} \right).
\end{split}
\end{eqnarray}
These formulas for covariances are the same in the case of the risk-neutral process (\ref{referenceportfolio_eq}) under the probability measure $\mathbb{Q}$.

\subsection{Bond and Vanilla prices}\label{bond_vanilla_price_appendix}
Zero coupon bond price $P(t,T)=\mathrm{E}_t^{\mathbb{Q}}\left[e^{-\int_{t}^{T}r(u)du} \right]$ in the case of Vasicek interest rate model (\ref{referenceportfolio_eq})  can be calculated directly using distribution of random variable $Y=\int_{t}^{T}r(u)du$ which is Normal with the mean and variance derived in Appendix \ref{distribution_Yt_sec}. This gives
\begin{eqnarray}\label{bond_price_eq_appendix}
\begin{split}
P(t,T)&=e^{-\mathrm{mean}(Y)+\frac{1}{2}\mathrm{var}(Y) }=e^{A_{t,T}-r(t)B_{t,T}},\\
B_{t,T}&=\frac{1}{\kappa}\left(1-e^{-\kappa (T-t)}  \right),\\
A_{t,T}&= -\kappa\int_{t}^T \theta B_{s,T}ds + \frac{\sigma_r^2}{4\kappa^3}\left( 2\kappa(T-t) + 4e^{-\kappa(T-t)}-e^{-2\kappa(T-t)}-3\right).
\end{split}
\end{eqnarray}
Allowing $\theta:=\theta(t)$ to be time dependent parameter, one can find $\theta(t)$ yielding bond prices observed at $t=0$. In the case of constant $\theta$, the above formula for $A_{t,T}$ simplifies to
$$
A_{t,T}=\left(\theta-\frac{\sigma_r^2}{2\kappa^2}\right)\left(B_{t,T}-(T-t)\right)-\frac{\sigma_r^2}{4\kappa}B_{t,T}^2.
$$

In the case of the risk-neutral process (\ref{referenceportfolio_eq}) for $S(t)$ and $r(t)$ under the measure ${\mathbb{Q}}$,
the closed-form formulas for prices of vanilla call $Q_{t_0}^{\mathrm{call}}$ and vanilla put $Q_{t_0}^{\mathrm{put}}$ with strike $K_T$ at maturity $T$, given $S(0)$ and $r(0)$, at time $t_0=0$ can be easily found. Changing measure ${\mathbb{Q}}$ to $\widetilde{\mathbb{Q}}$ and using formulas for the mean and variance of Normally distributed $\ln S(T)$ under $\widetilde{\mathbb{Q}}$ derived in Appendix \ref{appendix_Spdf}, obtain after simple calculus
\begin{equation}\label{vanilla_closedform_solution}
\begin{split}
Q_{t_0}^{\mathrm{call}}&=\mathrm{E}_{t_0}^{\mathbb{Q}}\left[e^{-\int_0^T r(u)du}\max(S(T)-K_T,0)\right]\\
&=P(0,T)\mathrm{E}_{t_0}^{\widetilde{\mathbb{Q}}}\left[\max(S(T)-K_T,0)\right]\\
&= S(0) N(d_1)- K_T P(0,T) N(d_2),\\
Q_{t_0}^{\mathrm{put}}&=\mathrm{E}_{t_0}^{\mathbb{Q}}\left[e^{-\int_0^T r(u)du}\max(S(T)-K_T,0)\right]\\
&=P(0,T)\mathrm{E}_{t_0}^{\widetilde{\mathbb{Q}}}\left[\max(S(T)-K_T,0)\right]\\
&= K_T P(0,T) N(-d_2) - S(0) N(-d_1).
\end{split}
\end{equation}
Here, $N(x)$ is the standard Normal distribution function, $P(0,T)$ is the bond price given by (\ref{bond_price_eq}), $d_1=(\ln({S(0)}/{K_T})-\ln P(0,T)+\frac{1}{2}\sigma^2_{\mathrm{eff}})/\sigma_{\mathrm{eff}}$, $d_2=d_1-\sigma_{\mathrm{eff}}$ and
\begin{equation}
\sigma^2_{\mathrm{eff}}=\sigma_S^2 T +\frac{\sigma_r^2}{2\kappa^3}\left(2\kappa T-3+4e^{-\kappa T}-e^{-2\kappa T}   \right)+\frac{2\rho\sigma_S\sigma_r}{\kappa^2}\left(\kappa T-1+e^{-\kappa T}\right).
\end{equation}
Note, the derived formulas are the same as the well known Black-Scholes formulas for call and put under the constant interest rate $r$ and volatility $\sigma$ if $rT$ is replaced by $-\ln P(0,T)$ and $\sigma^2 T$ is replaced by $\sigma^2_{\mathrm{eff}}$.

\renewcommand{\baselinestretch}{1.0}
{\small{
\bibliographystyle{chicago} 
\bibliography{bibliography}
}}

\end{document}